\documentclass[a4paper,fleqn]{cas-dc}

\usepackage{algorithm}
\usepackage{algpseudocode}

\usepackage[numbers]{natbib}

\def\tsc#1{\csdef{#1}{\textsc{\lowercase{#1}}\xspace}}
\tsc{WGM}
\tsc{QE}
\tsc{EP}
\tsc{PMS}
\tsc{BEC}
\tsc{DE}

\newcommand{\Exp}{\mathbb{E}}
\newcommand{\Cov}{\text{Cov}}
\newcommand{\Var}{\text{Var}}
\newcommand{\mb}[2]{#1^{(#2)}}
\newcommand{\ageto}[2]{#1\text{-}#2}

\begin{document}
\let\WriteBookmarks\relax
\def\floatpagepagefraction{1}
\def\textpagefraction{.001}

\shorttitle{Investigating HPV Dynamics With Emulation}

\shortauthors{A Iskauskas et~al.}

\title [mode = title]{Investigating Complex HPV Dynamics Using Emulation and History Matching}

%
\author[1]{Andrew Iskauskas}[type=editor,
                        orcid=0000-0003-2825-3651]

\cormark[1]

\ead{andrew.iskauskas@durham.ac.uk}


\affiliation[1]{organization={Department of Mathematical Sciences},
    addressline={Durham University}, 
    city={Durham},
    postcode={DH1 3LE}, 
    country={UK}}

\author[2]{Jamie A. Cohen}[orcid = 0000-0002-8479-1860]

\author[3]{Danny Scarponi}[orcid = 0000-0002-7587-9182]

\author[1]{Ian Vernon}[orcid = 0000-0002-9161-9946]
\author[1]{Michael Goldstein}[orcid = 0000-0002-0216-9913]
\author[2]{Daniel Klein}[orcid = 0000-0003-0437-7916]
\author[3]{Richard G. White}[orcid = 0000-0003-4410-6635]
\author[3]{Nicky McCreesh}[orcid = 0000-0003-1409-8531]

\affiliation[2]{organization={Institute for Disease Modelling},
    addressline={Bill and Melinda Gates Foundation}, 
    city={Seattle},
    country={USA}}

\affiliation[3]{organization={London School of Hygiene and Tropical Medicine},
city = {London},
country = {UK}}

\cortext[cor1]{Principal corresponding author}

\begin{abstract}
The study of transmission and progression of human papillomavirus (HPV) is crucial for understanding the incidence of cervical cancers, and has been identified as a priority worldwide. The complexity of the disease necessitates a detailed model of HPV transmission and its progression to cancer; to infer properties of the above we require a careful process that can match to imperfect or incomplete observational data. In this paper, we describe the \texttt{HPVsim} simulator to satisfy the former requirement; to satisfy the latter we couple this stochastic simulator to a process of emulation and history matching using the R package \texttt{hmer}. With these tools, we are able to obtain a comprehensive collection of parameter combinations that could give rise to observed cancer data, and explore the implications of the variability of these parameter sets as it relates to future health interventions.

\end{abstract}


\maketitle

\section{Introduction}

Human papillomavirus (HPV) is the most common sexually-transmitted infection, and is the dominant contributor to the incidence of cervical cancer globally \cite{burd2003human}. In recent years, the World Health Organisation (WHO) has identified HPV as a priority and introduced the Global Strategy to Accelerate the Elimination of Cervical Cancer \cite{brisson2019global}. In particular, attention has been focused on the delivery of effective one-dose vaccines \cite{who2022vaccine} that could make the elimination of HPV more attainable for many countries, not least those with the highest burden of cervical cancer.

To evaluate the benefits and drawbacks of different vaccine roll-out schemes in a particular country or region, without intensifying screening processes, we must have an understanding of the transmission, dynamics, and progression of HPV within the population of interest. A common approach to making such determinations is via a complex computer model -- an approach widely used in epidemiology, including for modelling COVID-19 \citep{krauss2022june}, Human Immunodeficiency Virus (HIV) \citep{andrianakis2017efficient} and tuberculosis \citep{scarponi2021tbhiv}. To this end, the \texttt{HPVsim} model \citep{stuart2023hpvsim} has been developed. This agent-based model allows for flexible and detailed simulations of HPV transmission through a population network using structured sexual networks, co-transmitting HPV genotypes, B- and T-cell mediated immunity, and high-resolution disease natural history. The model can also be used to apply interventions, such as the rollout of different vaccination strategies, to represent possible future policy suggestions and assess their potential impact and cost-effectiveness.

The detail of \texttt{HPVsim} allows it to accurately reflect the intricate dynamics of HPV, but comes at a cost. Before intervention strategies can be evaluated, we must first determine the combinations of parameters within the simulator that can give rise to the observed reality of HPV and cervical cancer epidemiology in the country or region in question. The high fidelity and depth of the model translates to a multitude of different parameters and choices of network structure and, without a systematic methodology for assessing the `suitability' of a given collection of parameters, any subsequent inference or prediction will be imperfect or incorrect. It is also necessary to accept that neither the model nor the real-world observations are expected to be perfect representations of the underlying reality of HPV progression and the true HPV and cervical cancer burden, respectively; any process of matching parameter sets to observations must account for this.

The structure of an agent-based model gives rise to an additional consideration -- even for a fixed choice of parameters the simulator is inherently stochastic, and repeated evaluations will result in a different prediction for the quantities of interest. While we could deem the variation due to stochasticity as subdominant to other effects, if we wish to find a complete collection of viable parameter choices upon which predictions and policy can be built then we should consider the stochastic behaviour as integral to the simulator output.

Various methodologies exist to match complex models to data \citep{akiba2019optuna, panovska2023machine, gibson1998estimating, o1999bayesian, jewell2009bayesian, mckinley2009inference, mckinley2018approximate, toni2009approximate, kennedy2001bayesian}, each with their own advantages and drawbacks. We choose here to apply the \emph{history matching} framework \citep{craig1997pressure}, which allows for an inherent understanding of model and observational uncertainty. Suitably applied, it can be used to find the complete space of parameter combinations that can give rise to observational data, in contrast to approaches like optimisation which seek to find a single `best' match. Using the history matching framework in conjunction with \emph{emulators} \citep{vernon2014galaxy} -- fast statistical approximations to model output -- we may explore the high-dimensional parameter space in a computationally feasible manner.

\section{The \texttt{HPVsim} Model}\label{sec:HPVsim}
Human Papillomavirus (HPV) is one of the most common sexually transmitted infections worldwide, with an estimated $80\%$ of people infected with it at least once in their lifetime \cite{chesson2014estimated}. Most infections are believed to clear of their own accord within $2$ years of onset \cite{adebamowo2018clearance}, but some infections persist and cause cells to transform into pre-cancerous lesions. Unchecked, these cells can develop into invasive cancers, particularly vaginal, anal, penile, and cervical cancers. Of real concern is the link between HPV and cervical cancer, as it is estimated that over $99\%$ of cervical cancer cases present with some form of HPV infection \cite{walboomers1999human}, contributing to over half a million new cases annually.

Despite the clear correlation between HPV and cervical cancers, the progression of pre-cancerous cells transforming to stages of cervical intraepithelial neoplasia (CIN) and finally to cancer can not be reliably observed due to long dwelltimes between infection, onset of pre-cancer, and invasion; the influence of HPV on this transition is therefore not necessarily straightforward. It is therefore necessary to use complex computer models (henceforth referred to as \emph{simulators}) to represent the natural history of the disease, in order to obtain estimates of cancers caused by HPV. Such simulators are more accessible -- and, in some cases, more reliable -- than observational studies of the disease; when used judiciously they can be a powerful tool to use to inform decision making for future policy.

One such simulator is \texttt{HPVsim} \cite{stuart2023hpvsim}, built under the umbrella framework of the Starsim models \cite{kerr2021covasim}. It allows for a detailed agent-based representation of the population dynamics of a country without requiring that each individual be modelled explicitly; gives an appropriate and flexible contact structure for HPV transmission; and allows a detailed specification of parameters that influence the natural history of the disease. Different HPV genotypes can be modelled separately -- a crucial aspect of modelling the disease, where different genotypes have been observed to contribute at varying severity to pre-cancerous lesions, CIN stages, and cancers \cite{hpvcentreNIG}. Intervention strategies, including adolescent screening and prophylactic vaccination, can be easily implemented; the process of natural clearance and reactivation of HPV within an individual can also be included. The combination of features within the model makes \texttt{HPVsim} flexible enough to accurately represent the dynamics of the disease and its progression to cancer within national or sub-national settings.

\begin{figure}[!ht]
    \centering
    \includegraphics[width = \columnwidth]{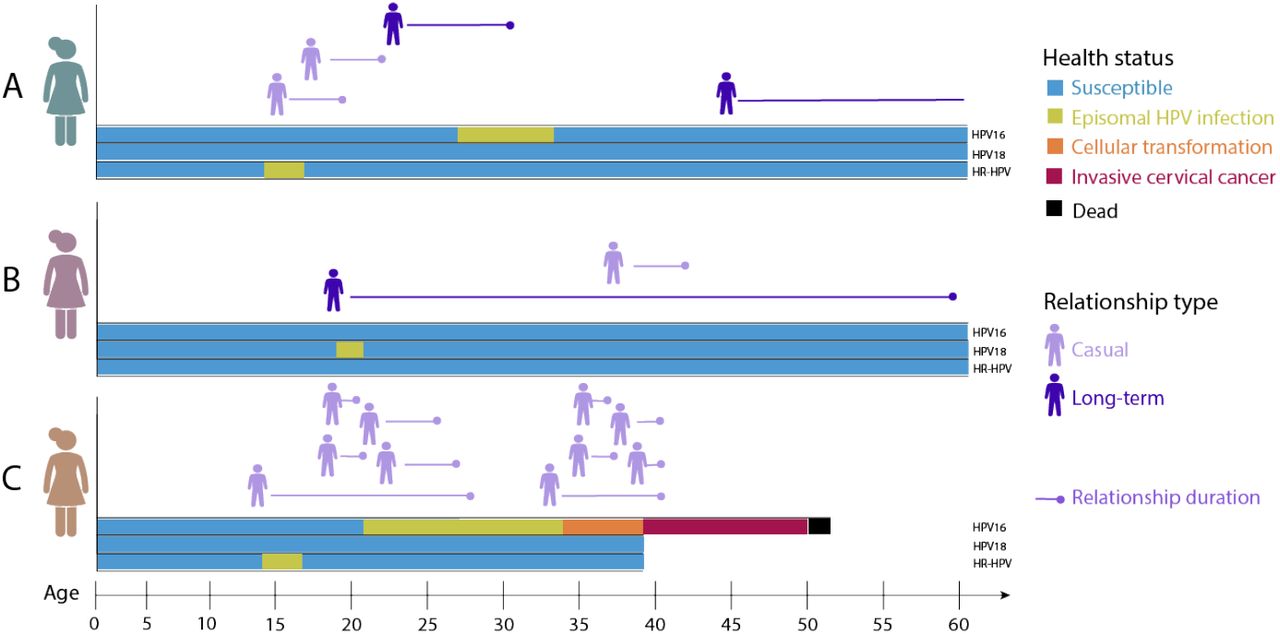}
    \caption{\small{Three possible individuals and their route through the HPV natural history (a detailed description of each route can be found in \cite{stuart2023hpvsim}).}}
    \label{fig:HPVpathway}
\end{figure}

An in-depth description of the structure of \texttt{HPVsim} \cite{stuart2023hpvsim} and a technical description of its usage \cite{hpvsimdoc2023} are beyond the scope of this article; we briefly describe the high level structure here. Individuals are classified by a sexual contact structure, which accounts for individuals partaking in marital, casual, and one-off relationships. Each of these classes of relationship has distinguishing characteristics such as average duration, probability of concurrency, and probability of condom usage, all of which may contribute to the likelihood of contracting HPV. Once HPV has been contracted by an individual, \texttt{HPVsim} models the corresponding natural history dependent on the genotype contracted and agent characteristics, such as prior immunity; in this, the infection has a probability of transforming cells and an individual may progress from pre-cancerous to having neoplastic and finally cancerous cells. An individual may clear the infection without external intervention; in this case, cells already transformed may remain and the individual may contract HPV again in their lifetime but with lower probability, depending on parameter choices regarding initial and waning immunity and cross-genotype immunities. Some examples of pathways that may occur within an individual in the \texttt{HPVsim} model are shown in Figure~\ref{fig:HPVpathway}, showing relationships and disease progression for three individuals within the model.

To avoid simulating each individual in the population of the country in question, and given the comparatively high prevalence of HPV within a community, a dynamic rescaling of the population is performed at the point where an individual joins the cancer pathway; one individual comes to represent a number of individuals with (initially) the same characteristics, all of whom then progress through the natural history dynamics at their own rate. This multiscale approach elegantly balances the needs of individual based modelling and the computational demands of running large populations, allowing for high granularity of results with efficient simulator run-time.

The flexibility of such a model coupled with the relative paucity of reliable data on HPV prevalence or progression to cancer, particularly in low- to middle-income countries (LMICs), poses a problem of identifiability. If we wish to find a collection of parameters that, when provided to \texttt{HPVsim} to produce simulated outputs, would give rise to the observed reality, the dimensionality of the input parameter space can vastly outstrip that of the available data. It therefore becomes crucial that we have a good understanding of parameters that materially affect the disease dynamics. Furthermore, even if we can identify a subset of the parameters at our disposal to investigate, there remain two questions to answer:
\begin{itemize}
    \item How can we best explore the (reduced) parameter space?
    \item By considering only a subset of parameters, are we removing parts of parameter space that could in fact represent reality?
\end{itemize}
The former question is common in the modelling community, particularly since many methods of calibration suffer from what is referred to as `the curse of dimensionality' \cite{fernandez2020curse} which might preclude a computationally viable in-depth exploration of the parameter space we wish to consider. The latter question is more subtle, but of paramount importance if we wish to perform post-hoc analysis on the model once matched to data (for example, considering interventions) -- if we do not or can not consider the effect of having fixed parameters that could have been varied, we may have ignored parameter combinations that represent the true dynamics of the disease. Any further analysis would therefore be at risk of misrepresenting the projected effects of our intervention strategies and result in non-robust inference: this could be devastating in the context of national health strategies.

One further complication is that \texttt{HPVsim} is inherently stochastic; if we wish to robustly explore the parameter space we must be aware that multiple evaluations, termed \emph{repetitions} or \emph{realisations}, will be required at each parameter combination to understand this source of variability. For complex models such as \texttt{HPVsim}, obtaining even a modest number of realisations at each parameter combination can be time- and resource-consuming and compounds the issue of exploring the space fully. We therefore apply a methodology capable of accounting for the two questions posed while handling the disconnect between the finite collection of realisations and the observed reality, in a computationally feasible time.

\section{Emulation and History Matching}\label{sec:hme}

In this section we briefly outline the mathematical framework of emulation, the process of history matching, and the effect of stochasticity on this approach. More details on the background to emulation can be found in \cite{craig1997pressure, vernon2018bayesian, santner2003design, bowman2016emulation}. Details about the framework, including the specific application of emulation and history matching to stochastic models, are provided in Appendix A.

An emulator \cite{vernon2014galaxy, jackson2019bayes, goldstein2013assessing} is a statistical surrogate to a complex simulator output, which given a relatively small ensemble of runs from the simulator in question can provide predictions of the simulator output at any unseen point in the parameter space. In this application, we construct \emph{Bayes linear} emulators \cite{goldstein2007bayes}, whose prior specification requires only a statement about expectations, variances, and covariances; given these specifications and data $D$ from simulator evaluations, an emulator can provide posterior predictions, $\Exp_D[f(x)]$ for the output $f(x)$ at any unseen point $x$, as well as the uncertainty in the prediction $\Var_D[f(x)]$. Computing the adjusted expectation and variance corresponds simply to matrix multiplication and a single offline computation of a matrix inverse, making their evaluation orders of magnitude faster than simulator evaluations from any but the most simple models.

Having created emulators for the outputs of interest, our aim is to leverage their computational efficiency to explore the parameter space and identify potentially suitable points for matching to observational data. We must first decide what it means for a point to be `suitable' in this context; to this end, we define an \emph{implausibility measure} $I(x)$:

\begin{equation}\label{eq:impmeasure}
    I^2(x) = \frac{(\Exp_D[g(x)]-z)^2}{\Var_D[g(x)]+\Var[e]+\Var[\epsilon(x)]}.
\end{equation}

Here, the random quantities $e$ and $\epsilon(x)$ represent the observational uncertainty and the model discrepancy respectively; to wit, the extent to which we are uncertain about the accuracy of the real-world observation, $z$, made and the (possibly $x$-dependent) extent to which we believe our simulator may not be representative of real life \cite{goldstein2013assessing}. A large value of $I(x)$ suggests that, despite any uncertainties inherent in the model or in the emulation at the parameter combination $x$, the prediction is extremely unlikely to give rise to an acceptable match to observational data. Conversely, a small implausibility at $x$ can be due to two main reasons: either the prediction is close to the observation, suggesting an acceptable match to data; or the uncertainties (particularly the emulator uncertainty $\Var_D[g(x)]$) are large at the point, suggesting a region of parameter space worthy of further investigation. Such points are termed \emph{non-implausible}.

The process of history matching leverages the concept of implausibility, and the fact that emulators require relatively few simulator evaluations to train, as follows. At each `wave' $k$ of history matching, we use a collection of simulator evaluations drawn from our current non-implausible region (denoted $\mathcal{X}_k$) to train emulators to outputs of interest. These emulators are then used to calculate implausibility across the whole of $\mathcal{X}_k$, resulting in a smaller non-implausible space $\mathcal{X}_{k+1}$. This new non-implausible space is sampled from, forming the new ensemble of points on which to perform simulator evaluations. This process continues until we reach a pre-agreed stopping condition; at this point, our final non-implausible region represents all parameter combinations that could give rise to the observational data, given our uncertainties about the observational data and simulator itself. The formal statement of the history matching process, as well as extensions to considering multiple outputs and the effect of stochastic simulators upon the implausibility measure, can be found in Supplementary Material B.

The process of determining prior specifications for emulators of complex simulators is seldom intuitive, and performing robust sampling from geometrically non-trivial non-implausible regions is often difficult. In this work we have taken full advantage of the \texttt{hmer} package \cite{iskauskas2022CRAN}, which automates the process of emulator training and adjustment with respect to data via \texttt{emulator\_from\_data()}, validates trained emulators for suitability via \texttt{validation\_diagnostics()}, and carefully proposes space-filling designs from the non implausible region via \texttt{generate\_new\_design()}. It also allows for intuitive visualisation of multiple different features of the trained emulator predictions, the proposed non-implausible region, and evolving features of the input and output spaces across multiple waves. More details of the functionality of the \texttt{hmer} package may be found in \cite{iskauskas2022emulation}, as well as in the examples and vignettes included in the package itself; where visualisations have been generated using the package in this work, we quote the function directly.

\section{Results}\label{sec:results}

\subsection{Problem Specification}
 
The \texttt{HPVsim} model can provide a wealth of information about any feature of the disease, its spread through a population, and the progression to cancerous lesions within an individual. However, observational data available is not of the same breadth or depth, so we must determine what data is suitable for matching our simulator evaluations to. We consider as observations of interest the numbers of new cancer cases recorded in $2020$ in the country of interest, aggregated by age, as well as the measured distribution of HPV genotypes in the population for patients presenting with high-grade lesions (termed the ``CIN3'' state) and cancers \cite{hpvcancers, bruni2019human}. These data were deemed to be reliable and sufficient for providing a meaningful history match, particularly as the eventual goal of the analysis of \texttt{HPVsim} simulations includes an analysis of genotype acquisition within the population and future cancer cases. A common clinical differentiation between different HPV genotypes motivated a further focus and partial amalgamation of genotypes of interest: expert elicitation suggested that a dozen genotypes were of interest, accounting for over $95\%$ of HPV cases \cite{hpvcancers}. Two genotypes, HPV16 and HPV18, were considered distinct while five further genotypes (HPV 31, 33, 45, 52 and 58) were collected into a `high impact' class, HPVhi5; the remaining five genotypes of interest were collected into an `other high risk' class, HPVohr. With this specification of genotype classes, $22$ outputs of interest were identified for which reliable observational data was available, along with an appropriate understanding of their observational uncertainty. Details of this observational data, as well as any uncertainties therein, may be found in Supplementary Material C.2. Initial pilot simulations suggested a tension between reported early- and late-age cancers, compared to the possible output of \texttt{HPVsim} (and by extension, the disease dynamics); in the absence of strong beliefs regarding systematic bias in the data-collection, we deemed it appropriate to increase the observational uncertainty for these outputs. 
 
We must also determine the most important parameters to include in our analysis, from the large array of parameters available for adjustment in \texttt{HPVsim}. After expert consideration to highlight the most influential parameters to the simulator, and sensitivity analyses to determine those with the greatest impact on model behaviour, the candidate set of parameters was reduced to a collection of $33$ whose inclusion would be critical to a robust investigation of the model behaviour. For those parameters which remained unvaried, fixed values were determined via expert elicitation or inferred from demographic surveys \cite{statcompiler} and appropriate model discrepancy included to account for their exclusion. The initial non-implausible space, $\mathcal{X}_0$, therefore consisted of a $33$~dimensional hypercube whose extent corresponded to the largest physically reasonable range for each of the parameters considered. The initial parameter space is detailed in Supplementary Material C.1.

\subsection{Emulation and History Matching}

\subsubsection{Emulator Training}
At each wave, an ensemble of $495$ parameter sets in a space-filling design were provided to \texttt{HPVsim} for evaluation; of these, $330$ were used to train the emulators in accordance with arguments presented in \cite{loeppky2009choosing}, with the remaining $165$ used for validating the trained emulators. Due to the philosophy behind variance emulation and history matching, we were able to modify our emulation strategy throughout the waves without concerns about jeopardising our final inference. At early waves, anticipating volatile (and therefore computationally expensive) simulator behaviour, we demanded only $16$ realisations at each parameter combination and focused only on emulating the mean and variance response, accepting that this would result in a conservative prediction of uncertainty. As the non-implausible space was reduced and the simulator behaviour became more stable, we increased the number of realisations at a point to $32$, and at later waves still to $50$.

\begin{figure}[!ht]
    \centering
    \includegraphics[width=\columnwidth]{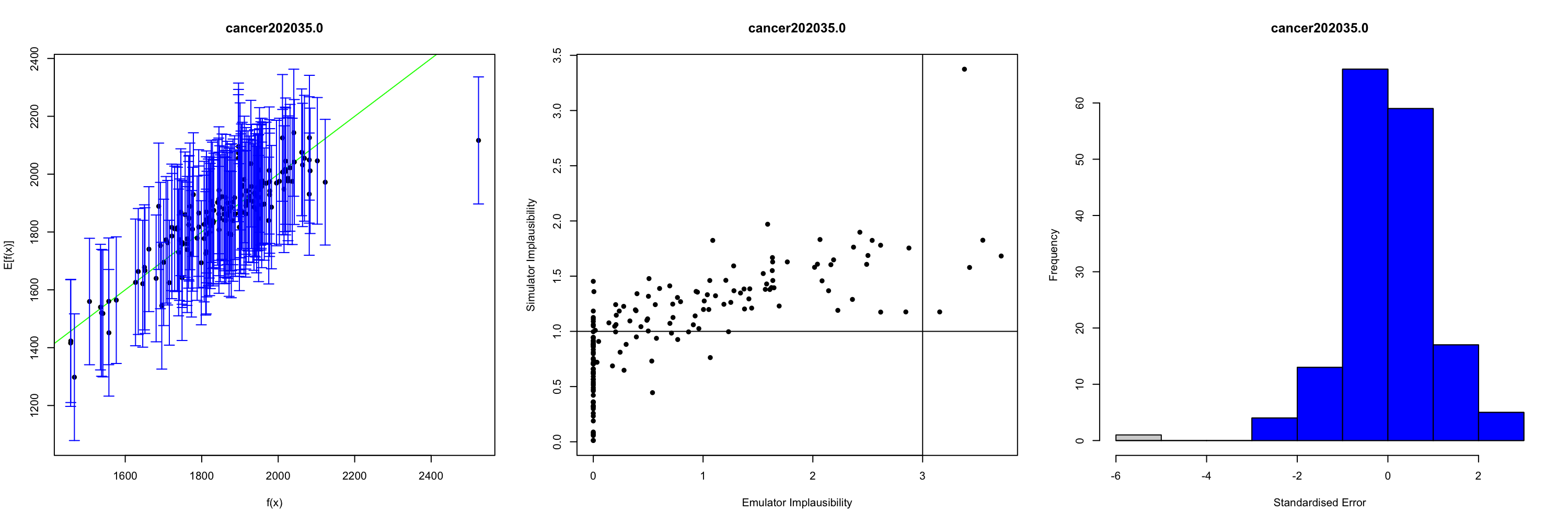}
    \caption{An example of diagnostics performed on each emulator. Details of each plot can be found in the \texttt{hmer} documentation.}
    \label{fig:diagnostics}
\end{figure}

To validate the specifications for our emulators, particularly the global response to the inputs, emulator diagnostics described in \cite{iskauskas2022emulation} were performed using the hold-out validation set, alongside a handful of `acceptable' simulator runs obtained via pilot studies and hand-fitting; where problems were observed in a particular emulated output, we were at liberty to interrogate the structure of the offending emulator and adjust the prior specification accordingly. An example of diagnostic plots produced for each emulator is shown in Figure~\ref{fig:diagnostics}. Corrective measures applied included inflation of the prior uncertainty; modification of the global emulator response; and inclusion of additional simulator runs in problematic regions of space. Quantities whose emulator could not adequately represent the output at a given wave, even after modification, were removed from the current wave and reincorporated at the successive wave.

\subsubsection{History Matching}\label{subsubsec:hmdesc}
Having created a collection of emulators for the relevant outputs at a wave, we applied the process of history matching. At early waves, we used a somewhat conservative second-maximum of univariate implausibilities, letting at most one output be far from the corresponding target, with a cutoff of $I_{2M}(x)\le3$; at later waves we required maximum implausibility to satisfy this cutoff. At the final wave, we incorporated relationships between emulated outputs using a multivariate measure. In searching for acceptable matches to observation, we leveraged the vastly improved computational efficiency of the emulators, evaluating no fewer than $10{,}000$ candidate points during the point proposal stage at each wave. We followed the procedure detailed in \cite{iskauskas2022emulation} to search for a space-filling non-implausible sample of points, summarised in Supplementary Material B.

Where we were unable to generate sufficiently many candidate points from this process, and hence we might be concerned about over-sampling of a small region of the true non-implausible region, the implausibility cutoff was allowed to be relaxed so as to allow for a representative sample. Candidate points generated in this `relaxed' framework were then sub-selected to provide a representative sample at a lower cutoff and used as a proposal for generating a larger collection of points. This process could be repeated many times before we obtained a space-filling design of non-implausible points at the desired cutoff, ensuring a more accurate identification of the complete non-implausible region\footnote{This `annealing' process is similar in spirit to that proposed in \cite{williamson2013implausibility}, though less computationally expensive and somewhat less robust in identifying disconnected regions. However, for this application, the method described above proved adequate for identifying the non-implausible space.}.

Our decision on how many waves were required for this problem was motivated by the aims of the analysis. There are two connected but different goals of the analysis at hand: the first is of interpretation of the non-implausible region, and in particular the most influential or restricted parameters; the second is to make meaningful statements about the distribution of different genotype classes within the population and the trend of future cancer cases, for which no observational data exists. To our first aim, we wished to propose sufficiently many points that, when evaluated by the simulator, gave rise to matches to observational data -- this was not guaranteed to occur and a secondary condition was that the emulator uncertainties had a subdominant effect in the denominator of the implausibility measure (suggesting that further waves of emulation would not materially affect the proposal of points). To address the second aim, we also evaluated a sample of the points proposed at each wave with respect to those unobservable quantities of interest. Where these quantities showed no substantial difference between waves or across the current non-implausible space, indicating that our further inference about these quantities would not be affected by a change in structure of the non-implausible region, we could consider this aim of the history match fulfilled.

\subsection{Analysis of Results}

\begin{figure}
    \centering
        \includegraphics[width=\columnwidth]{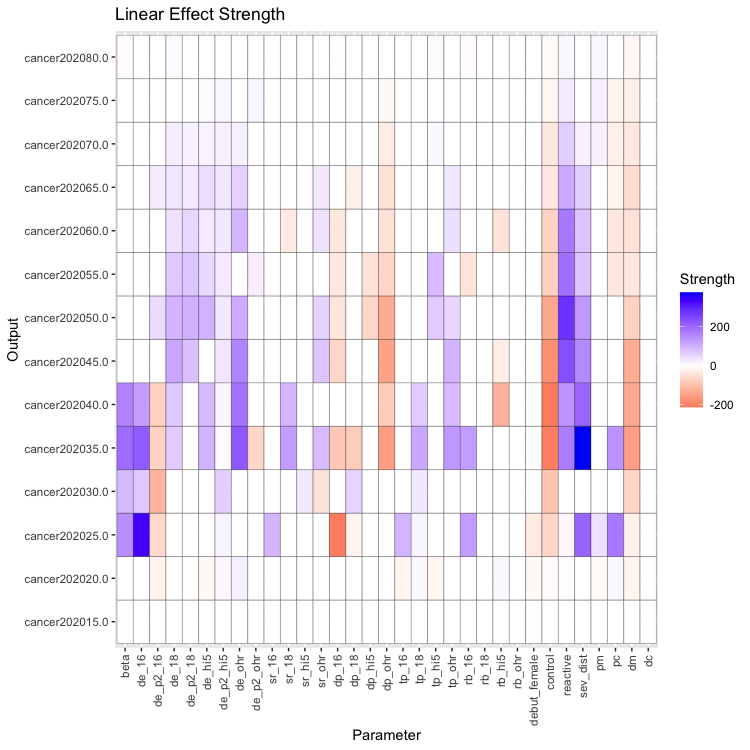}
    \caption{Active variables and strength of influence for late-wave emulators of cancer totals. Each row corresponds to an output of the model, and each column to a parameter. Stronger blue or red corresponds to a more significant positive or negative response, respectively.}
    \label{fig:effstrength}
\end{figure}

Before focusing on the eventual structure of the final non-implausible region, we first discuss some interpretable aspects of the simulator extracted from the emulation process. We can interrogate the active variables for each output and well as the strength of effect that each input has on the outputs in Figure~\ref{fig:effstrength}, as generated by \texttt{hmer::effect\_strength()}.

We may note some features of the outputs. Firstly, the choice of active variables coincides in many cases with physical intuition we might have about the disease: for example, transmission rate \texttt{beta} and episomal duration (the average duration of infection prior to clearance, control, or transformation) \texttt{de\_16} have a greater influence on the resulting early-age cancers than in later ages, while parameters governing latent control and reactivation of HPV (\texttt{control} and \texttt{reactive}) tend to more severely impact the resulting cancers in older age groups. The relative strengths-of-effect also agree with intuition: for example, a higher \texttt{beta} results in more cancer cases, while a higher \texttt{control} reduces the total number of cancers in an age group. Some aspects are more surprising: the relative paucity of impact that genotype-specific relative transmission rates \texttt{rb} have on cancers suggests that deviations from the overall transmissibility do not materially affect the resulting profile of cancers, perhaps suggesting that a combination of severity rate (the rate of transformation of cells in infected individuals) \texttt{sr}, relative transmissibility \texttt{rb}, and transformation probability \texttt{tp} may be more directly informative.

\begin{figure}[!ht]
    \centering
    \includegraphics[width=\columnwidth]{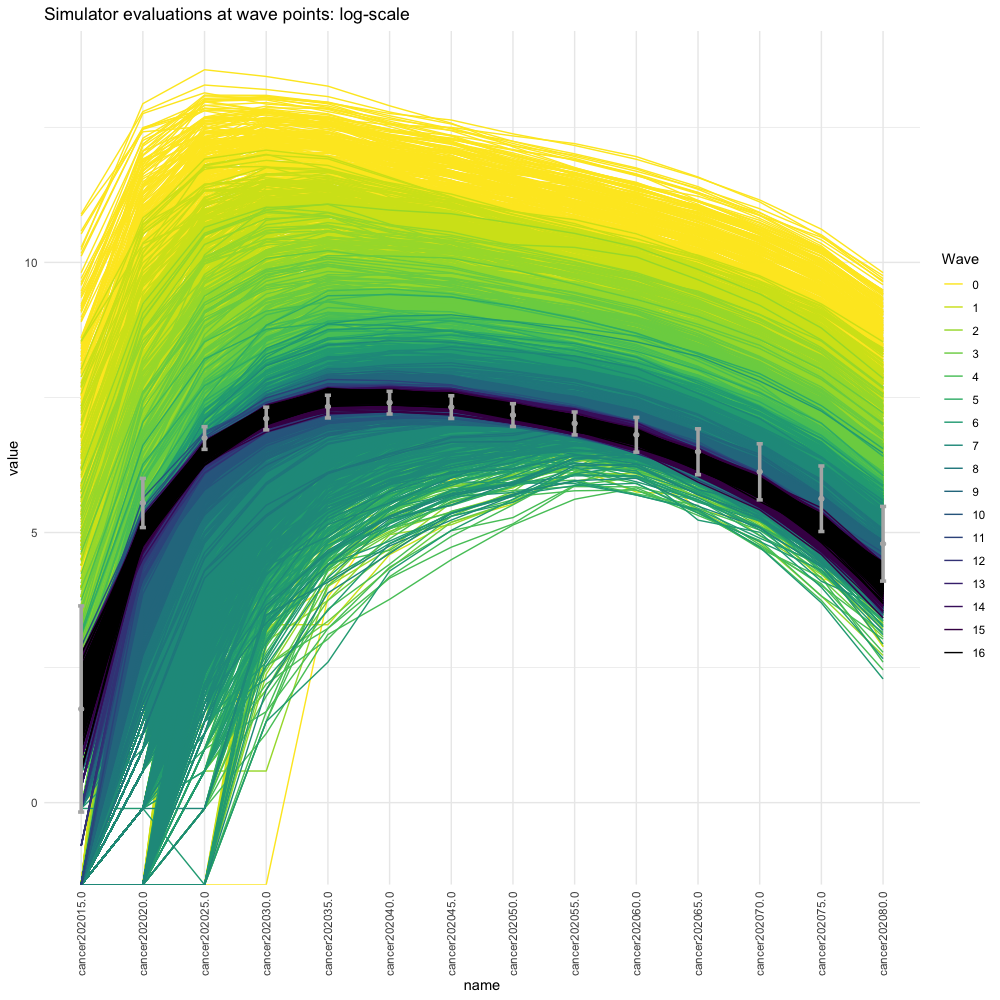}
    \caption{Plots of the model runs for cancer totals, coloured by wave, after $16$ waves of history matching. Totals are shown on the log-scale.}
    \label{fig:simcancer}
\end{figure}

\begin{figure}
    \centering
    \includegraphics[width=\columnwidth]{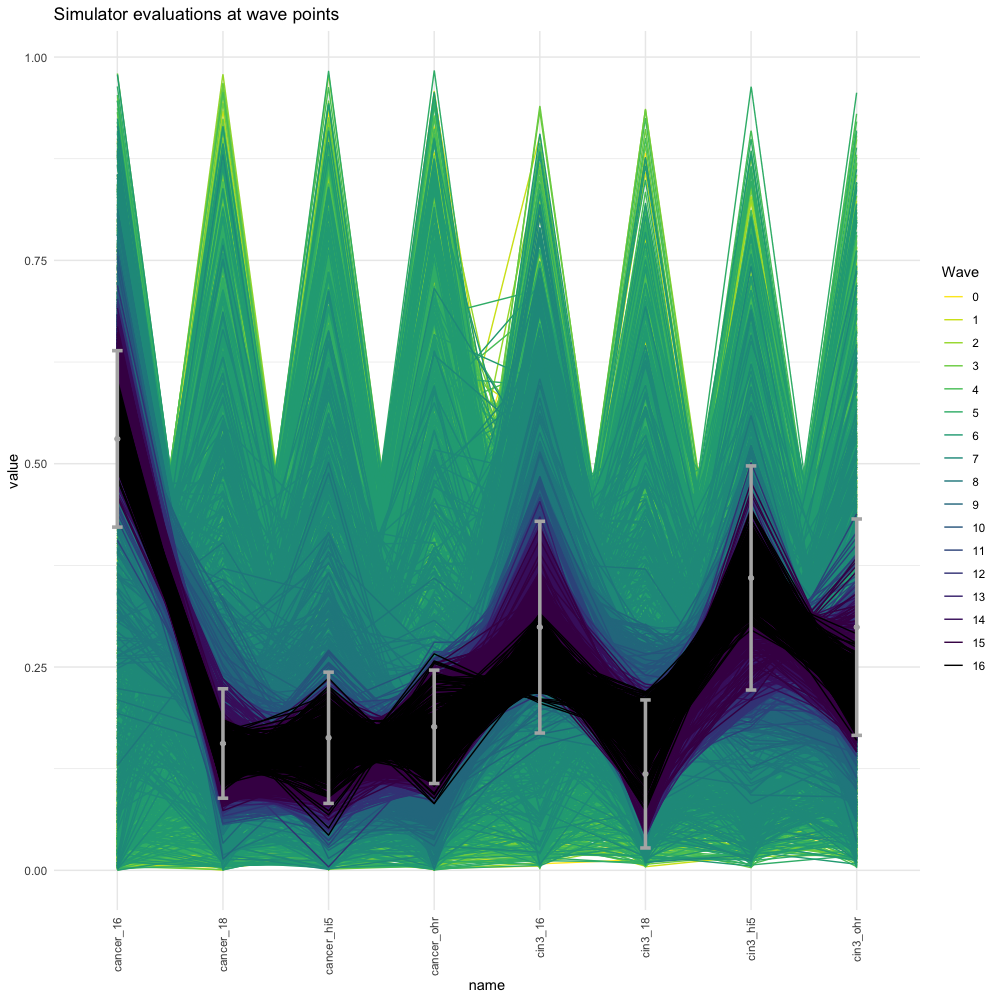}
    \caption{Plots of the model runs for genotype proportions contributing to cancers and high-grade lesions (cin3), coloured by wave, after $16$ waves of history matching.}
    \label{fig:simgeno}
\end{figure}

In all, considering the stopping conditions described above, $16$ waves of emulation and history matching were required. At this stage, the non-implausible region was unlikely to be restricted any further by additional waves, over $25\%$ of points proposed by the emulators resulted in complete matches to data (even in the absence of any model discrepancy), and all projected quantities of interest were stable between waves. To interpret the results most effectively, we now turn to visualisation of the final non-implausible space. We first consider the progression of the outputs towards the observational targets as we progress through the waves: Figures~\ref{fig:simcancer} and \ref{fig:simgeno} shows the output runs, coloured by wave, as generated by \texttt{hmer::simulator\_plot()}. For ease of visual analysis, we have split the outputs into two plots: one for cancer totals and one for proportions. Due to high output variability, the total number of cancers per age group is presented on a log scale.


With increasing wave number, the ensemble of simulator runs arising from non-implausible points at that wave is tightening toward the targets, shown as error bars. This reduction of the output space is particularly noticeable in Figure~\ref{fig:simcancer}, where the extrema of model behaviour in the first and final wave differs by a factor of up to $400$. The conflict anticipated in pilot studies is borne out by the early- and late-stage cancer totals, which are tied to the lower ends of their respective targets. Figure~\ref{fig:simgeno} demonstrates that there is more flexibility in the proportions of various HPV genotypes contributing to cancers and high-grade lesions, but the behaviour of genotype $16$ seems to be the most restricted beyond that imposed by the observational data. Note that in these plots, each individual line corresponds to a single parameter set, where we have aggregated using the mean of the realisations. There is therefore additional stochastic variability not displayed here, but this representation serves well in showing the progression of the history match and the broad trends of the simulator response.

\begin{figure}[!ht]
    \centering
    \includegraphics[width=\columnwidth]{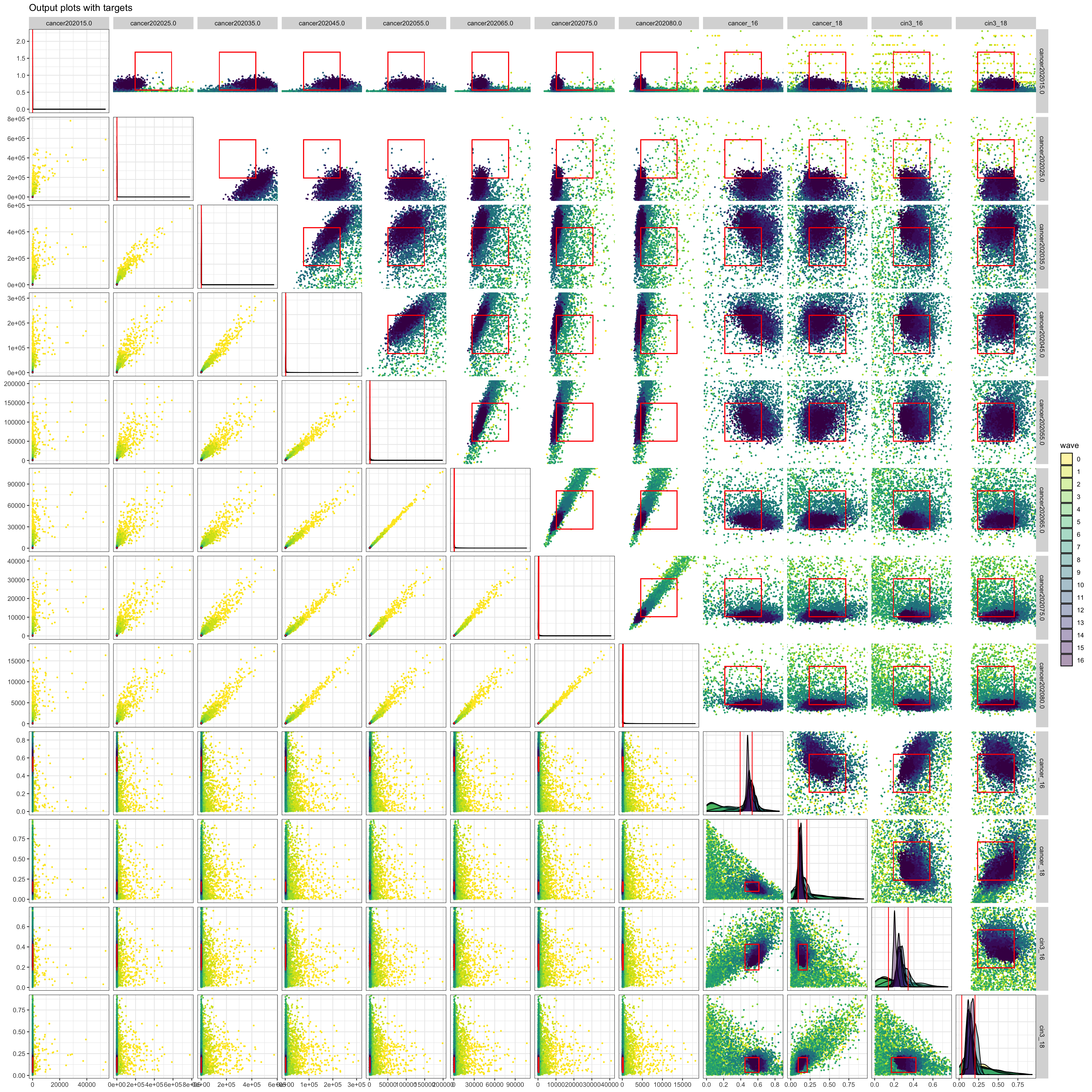}
    \caption{Simulator output plotted in pairs; each plot corresponds to two outputs, with a red rectangle overlaid representing the target region for those two observations.}
    \label{fig:wavevalues}
\end{figure}

To try to gain further insight into the output structure of the model, we might consider looking at pairs of observations rather than each individually. This information is presented in Figure~\ref{fig:wavevalues} for an informative subset of the outputs and provided using \texttt{hmer::wave\_values()} -- in each panel of the plot not on the main diagonal, the results of two simulator outputs for a given parameter set are plotted as a point, with a target window overlaid corresponding to the observational data. In those plots above the diagonal, the plot is `zoomed in' to the region of interest, so that the panel is centred on a square corresponding to the rescaled target window; below the main diagonal, the results are presented raw. We immediately see the extreme reduction in simulator behaviour as the history matching waves progress, to the extent that the observational targets for cancer cases are almost too small to see in the raw plots. We also see strong correlations between cancers in neighbouring age groups, attenuating as we consider correlations between age groups with larger separation. The conflict between early-age cancers and those at later age groups becomes more stark, as we see that simulator runs are only just able to overlap with the target windows in a small corner. This feature is especially stark when considering simultaneously matching to cancers at ages $\ageto{25}{30}$ and $\ageto{35}{40}$. In the targets where we consider proportions of genotypes, we see clear structure; particularly a hard diagonal bound on the value of pairs of outputs which deal with the same type of proportion (either cancers or CIN3), since these proportions must sum to no more than $1$; between different types of proportions, we see a broadly correlated relationship for a given genotype (for example, \texttt{cancer\_16} and \texttt{cin3\_16}), with little dependence between genotypes. This sheds light on the nature of the cross-genotype immunity structure we could have specified: were we to have had strong views on whether contracting and clearing HPV16 affords protection against contracting HPV18 (say) at some later point in life, we could have stated this within the fixed parameters of \texttt{HPVsim}; the results we have obtained here do not give any strong suggestion of the necessity of such a statement.

\begin{figure}[!ht]
    \centering
    \includegraphics[width=\columnwidth]{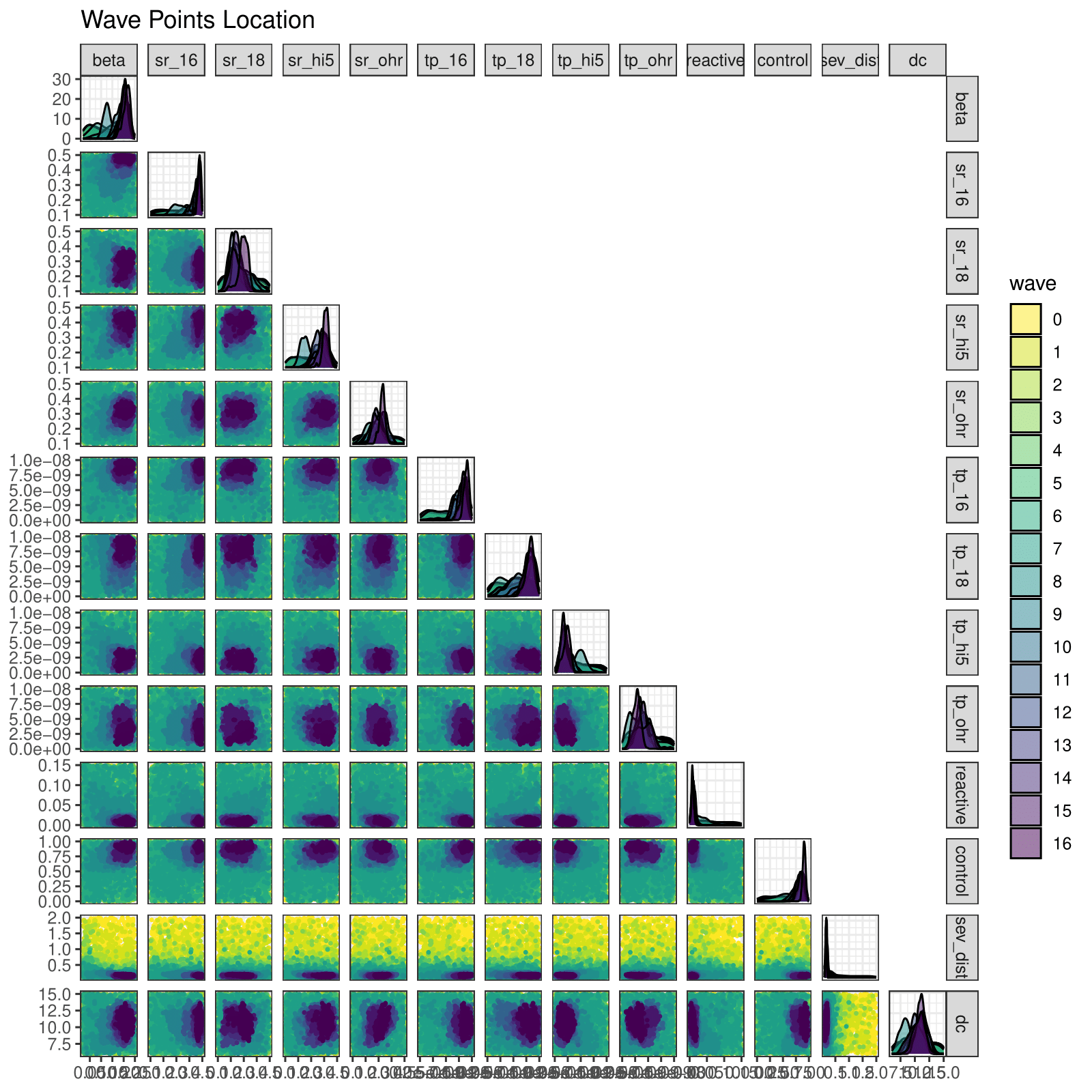}
    \caption{Input parameter sets plotted as pairs, coloured by wave.}
    \label{fig:wavepoints}
\end{figure}

We can examine two-dimensional projections of the input space, rather than the output space, using \texttt{hmer::wave\_points()}. In the interests of ease of inspection, a truncated collection of parameters is shown in Figure~\ref{fig:wavepoints} to highlight which input parameters have been most restricted across the waves of history matching, and to where.

We can see that, even at early waves where the simulator behaviour was subject to a large amount of volatility, the overall severity rate of the disease (governed by \texttt{sev\_dist}) was necessarily low. This is a consequence of the wide ranges placed on the parameters initially: while the highest values of severity were not unphysical, they were very unlikely to give rise to the observations we intended to match to and this was quickly borne out. We also see that the probability of reactivation of the disease and of latently controlling the disease (\texttt{reactive} and \texttt{control}, respectively) have found themselves heavily restricted by our final wave: in the absence of preventative measures and in light of observational data, we must conclude that the chance of reactivation of latently controlled HPV is low, and that a large number of infected individuals clear the disease of their own accord. This seems sensible in the context of HPV, where we know that an extremely high proportion of the population contract it, but comparatively few progress to cancer. It is possible, however, that introducing measures such as vaccination or screening into the \texttt{HPVsim} natural history would have modified these statements -- for this LMIC country in question, we would not consider this result to be incompatible with the true natural history of the disease due to limited testing and control strategies implemented. The genotype-specific parameters also show some interesting features: we see in particular that the individual genotypical severity rates show a large amount of dissimilarity with HPV16 having the largest severity of all genotypes under consideration; the other individual genotype, HPV18, seems to have one of the lowest severity profiles but its probability of transformation (\texttt{tp\_18}) is much higher than the aggregated genotypes HPVhi5 and HPVohr, partially justifying its singular inclusion.

We may find rough bounds on the volume of space that remains after performing the waves of history matching: on the basis of the proportion of non-implausible points in $\mathcal{X}_k$ that remain non-implausible according to the wave $k$ emulators, we estimate a final volume equal to $\sim 1\times10^{-16}$; a rough upper bound on the space reduction can be given by considering the minimum enclosing hyperrectangle for wave $16$ and compare it to our initial ranges, finding a volume ratio of $1.2\times10^{-14}$.

\begin{figure}[!ht]
\centering
    \includegraphics[width=\columnwidth]{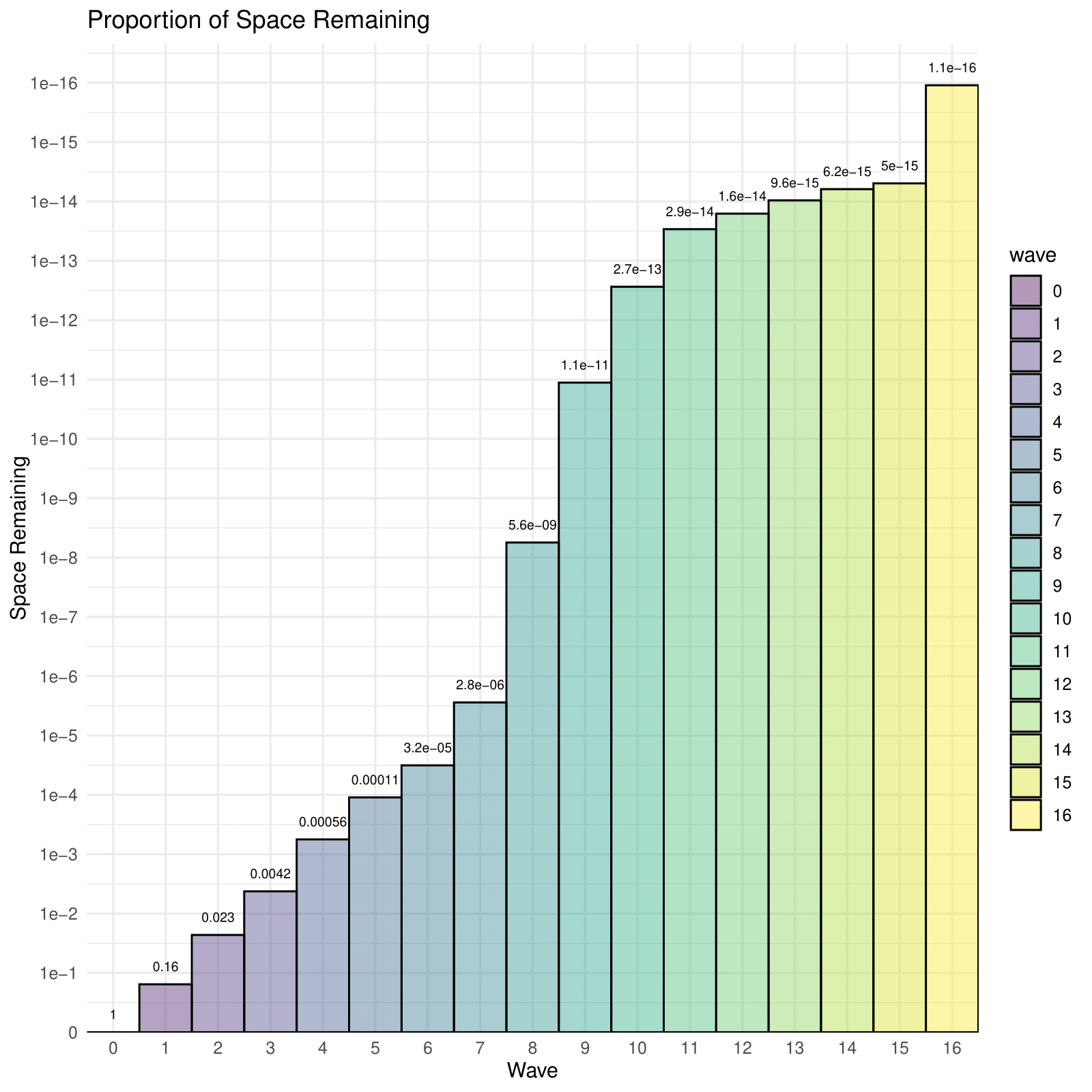}
    \caption{Ratios of space remaining per wave, relative to the initial parameter region $\mathcal{X}_0$.}
    \label{fig:timetaken}
\end{figure}

Figure~\ref{fig:timetaken} shows the reduction in space over the course of the waves of history matching. We may note the `spike' in space reduction in the final wave after a period of decreasing gains in previous waves: this is due to the inclusion of a multivariate implausibility measure in wave $16$, forcing relationships between outputs. In fact, the inclusion of covariance information between outputs was a large driver in the satisfaction of our stopping conditions; to verify this, a further ensemble of emulators was generated from the proposal at the end of wave $16$ but was unable to reduce the space to any significant extent. We also considered the emulator uncertainty at the final wave compared to the other sources of uncertainty, namely observational error and model discrepancy: the prior emulator variances are $4\%$ the size of the other uncertainties on average, with range $[0.2\%, 12.4\%]$. At this stage, the emulators are contributing very little to the overall uncertainty in the implausibility measure \eqref{eq:impmeasure}, and so further refinement of the emulators was extremely unlikely to lead to more accurate results. We return to the final stopping condition shortly.

The information presented in Figure~\ref{fig:timetaken} is worthy of further emphasis: to obtain a single point from the final non-implausible space were we to search at random from the initial volume, we would require around a quadrillion model evaluations (multiplied by the requisite number of realisations at each parameter set). The entire process of history matching and emulation required the evaluation in \texttt{HPVsim} of $8{,}000$ points in parameter space, including those that contributed to validation sets, resulting in $180{,}000$ model evaluations in total including repetitions. Of all points proposed, $802$ result in output consistent with all observed data - an appreciable yield of $10\%$. In the final wave of history matching, over half of the points proposed were consistent with observation and model discrepancy, and we may therefore use the emulators at the final wave to generate arbitrarily many non-implausible parameter sets at a speed far greater than would be possible using the simulator alone. The reason we have required so few simulator evaluations is due to the burden of computation and prediction taken on by the emulators. Only because emulator prediction of the mean response is around $10^7$ times faster than the equivalent \texttt{HPVsim} simulator evaluation, accounting for the need to perform realisations, were we able to effect a robust examination of the input parameter space.

\subsection{Comparison to Other Methods}
Of course, other methodologies exist to match complex models to data, and in fact \texttt{HPVsim} incorporates an optimisation algorithm \cite{akiba2019optuna}; we conclude this section with a comparison between our approach and that of the `native' \texttt{HPVsim} approach. 

\begin{figure}[!ht]
    \centering
    \includegraphics[width=\columnwidth]{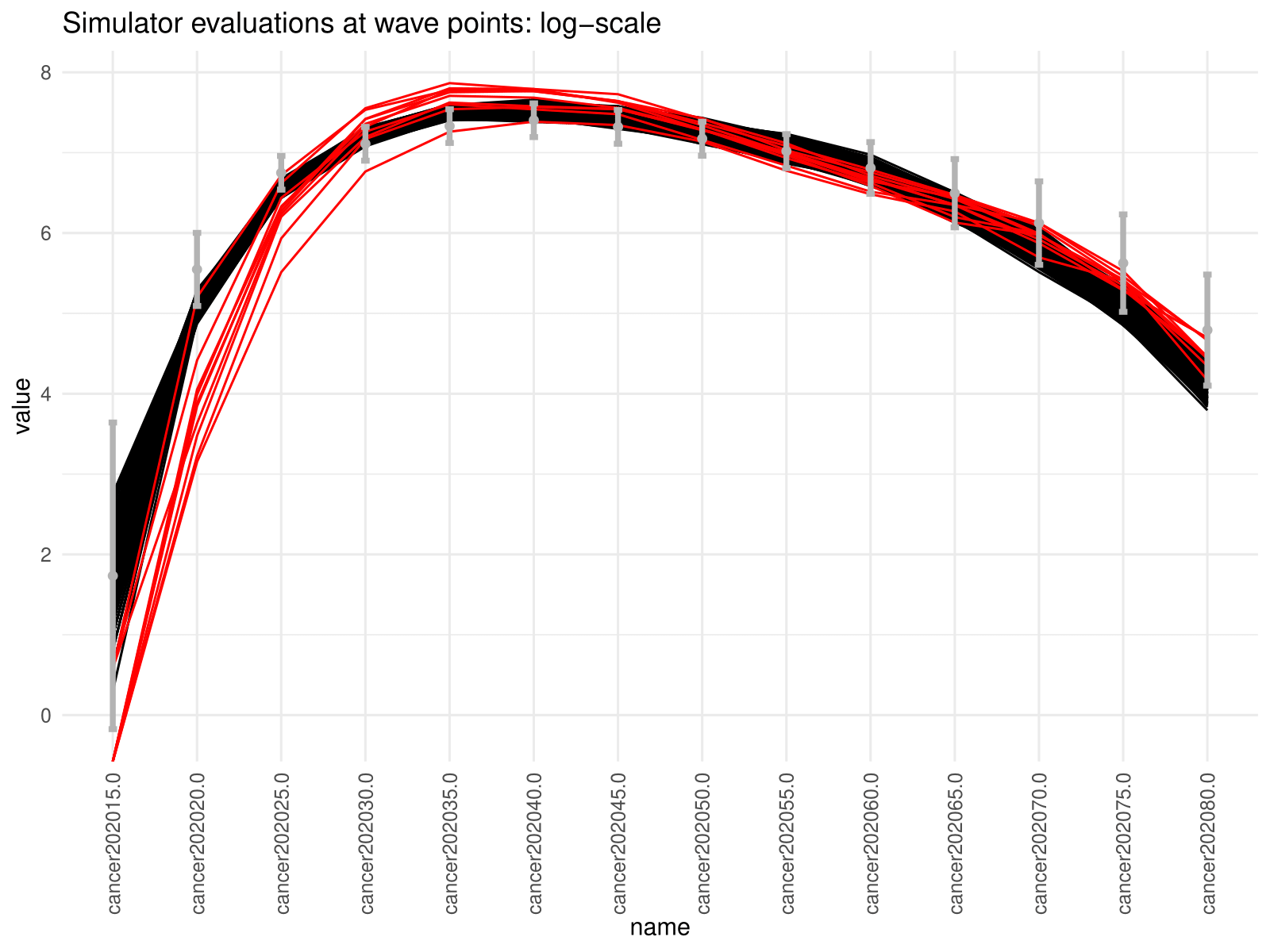}
    \caption{A comparison of cancer totals from model runs proposed by optimisation (red) and history matching (black).}
    \label{fig:emvoptcancer}
\end{figure}

\begin{figure}
    \centering
    \includegraphics[width=\columnwidth]{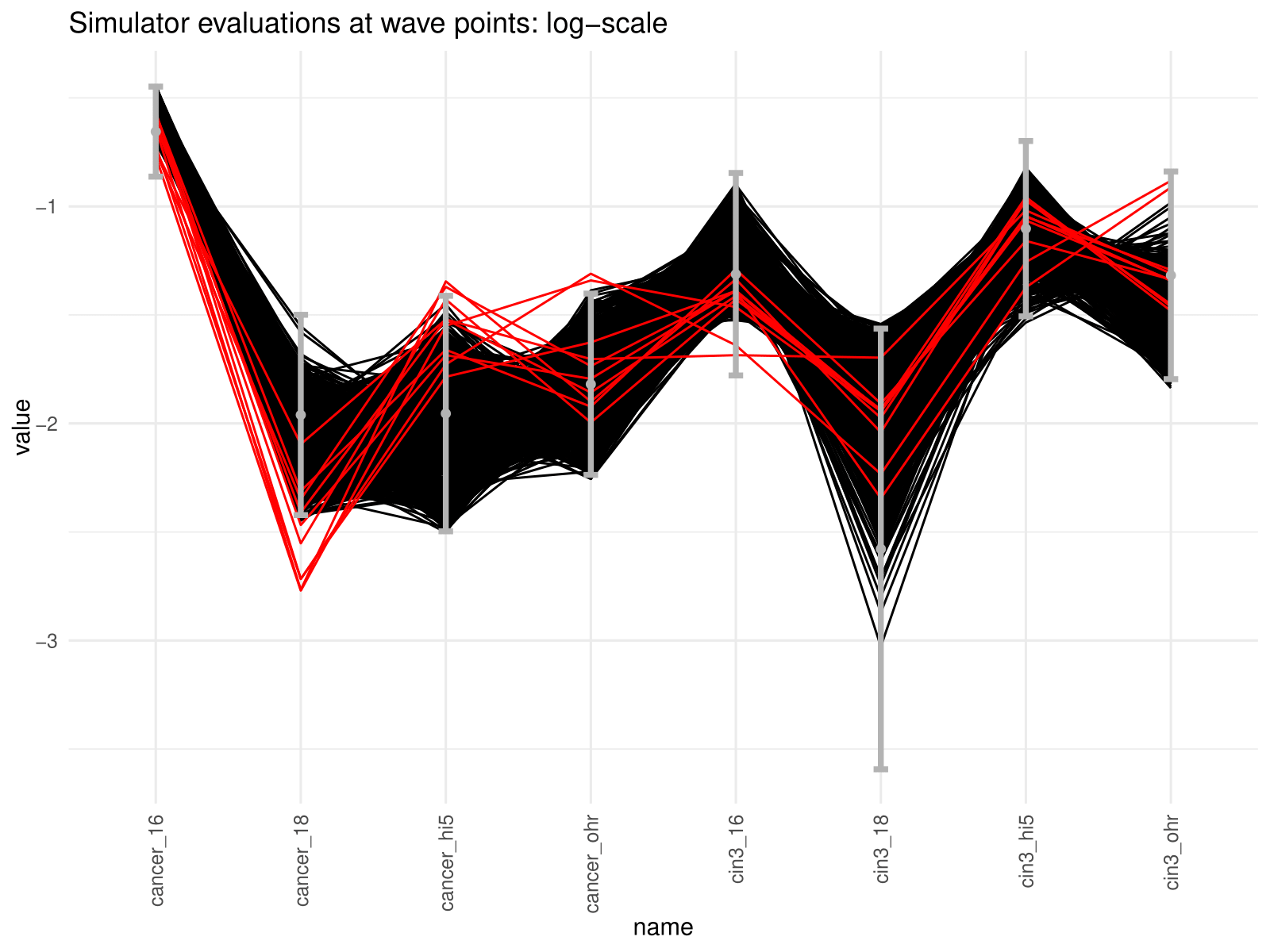}
    \caption{Comparison of HPV genotype proportions from model runs proposed by optimisation (red) and history matching (black).}
    \label{fig:emvoptgeno}
\end{figure}

Even with this comparatively inexpensive method, an optimisation run required between $1{,}000$ and $10{,}000$ simulator evaluations to find a local maxima of the goodness-of-fit measure, even before stochasticity is taken into account. Were we to assume that multiple seeds of the optimisation algorithm would result in an equivalent exploration of the non-implausible region, and supposing that we take the last $10$ points proposed by optimisation as a proxy for the space of interest, we would need to evaluate the simulator at a minimum of $80{,}000$ parameter locations; in order to use no more simulator evaluations than the history matching process required, we could therefore perform no more than $2$ realisations at each parameter combination, in contrast to the minimum of $16$ that we were able to evaluate with the history matching approach. Furthermore, the requirement of targeting a goodness-of-fit measure can provide proposals with good measure statistics that, under inspection, have unwanted behaviour when considered in the context of matching the model to data. We see this demonstrated in Figures~\ref{fig:emvoptcancer} and \ref{fig:emvoptgeno}, where the two methods are compared directly.


We can see that, when considering the age-aggregated cancers in Figure~\ref{fig:emvoptcancer}, those points proposed using history matching and those proposed via optimisation have somewhat different characteristics. Particularly for early-age groups, the behaviour of each group diverges: while the history matched proposals have been constrained to the lower ends of the target bounds in early- and late-age cancers so as to not overestimate mid-age cancer cases, the optimisation procedure (and the goodness-of-fit measure used) has prioritised targeting the centres of the late-age cancer cases, at the expense of those targets for cancers in anyone aged below $35$. In some cases, this disconnect is quite stark, even on a logarithmic scale (in particular, those for age groups $\ageto{20}{25}$ and $\ageto{25}{30}$). In fact, those points proposed by history matching surpassed the points generated via optimisation even with reference to the goodness-of-fit measure used by the optimiser, highlighting the risks inherent in using a method which can gravitate to local minima, even given different starting seeds.

In the targeted proportions shown in Figure~\ref{fig:emvoptgeno}, the difference is less striking when considering whether proposals lie inside the target window, but there is one aspect worthy of note. While the history matching proposals are spread widely around their respective target regions, those obtained via optimisation are comparatively tightly clustered and often to one extreme of the region that we know \emph{a posteriori} would result in acceptable matches. Were we to rely solely on the information provided to us by optimisation, we would be at risk of making overly restrictive statements about the relative contributions of each genotype to cancers, which could result in non-robust statements about the effectiveness of vaccination strategies were we to consider a single genotype in isolation. 

The above considerations, and particularly those arising from Figure~\ref{fig:emvoptgeno}, highlight the key conceptual difference between history matching and optimisation. Whereas optimisation seeks to find a `good' parameter combination for fitting to observational data, history matching seeks to find \emph{all} possible combinations that match observations. We would therefore not expect the optimisation algorithm to span the full space of possible matches (even with multiple restarts with different seeds), as it is task to which it is not designed. The primary strength of optimisation is the speed at which one might find `good' parameter combinations, but we see here that even this cannot be depended on: to stand a chance of achieving an collection of acceptable parameter combinations of comparable size, accounting for stochastic repetition, we would have required at least an order of magnitude more simulator evaluations. Furthermore, there is no guarantee that those points generated would be representative of the full space of good matches, liable as the algorithm is to settle in local minima of some loss function.

\section{Inference and Prediction for HPV}\label{sec:analysis}

Often, a key aim of disease modelling is to make informed statements about the future trend of the disease (as well as the possible spread of outcomes) so as to make robust recommendations about the potential efficacy of intervention campaigns; we may also wish to investigate properties of the natural history of the disease that would be impossible, or at least extremely hard, to do via real-world studies. We detail here how the results of the history matching process helps with both of these aims.

\begin{figure}[!ht]
    \centering
    \includegraphics[width=\columnwidth]{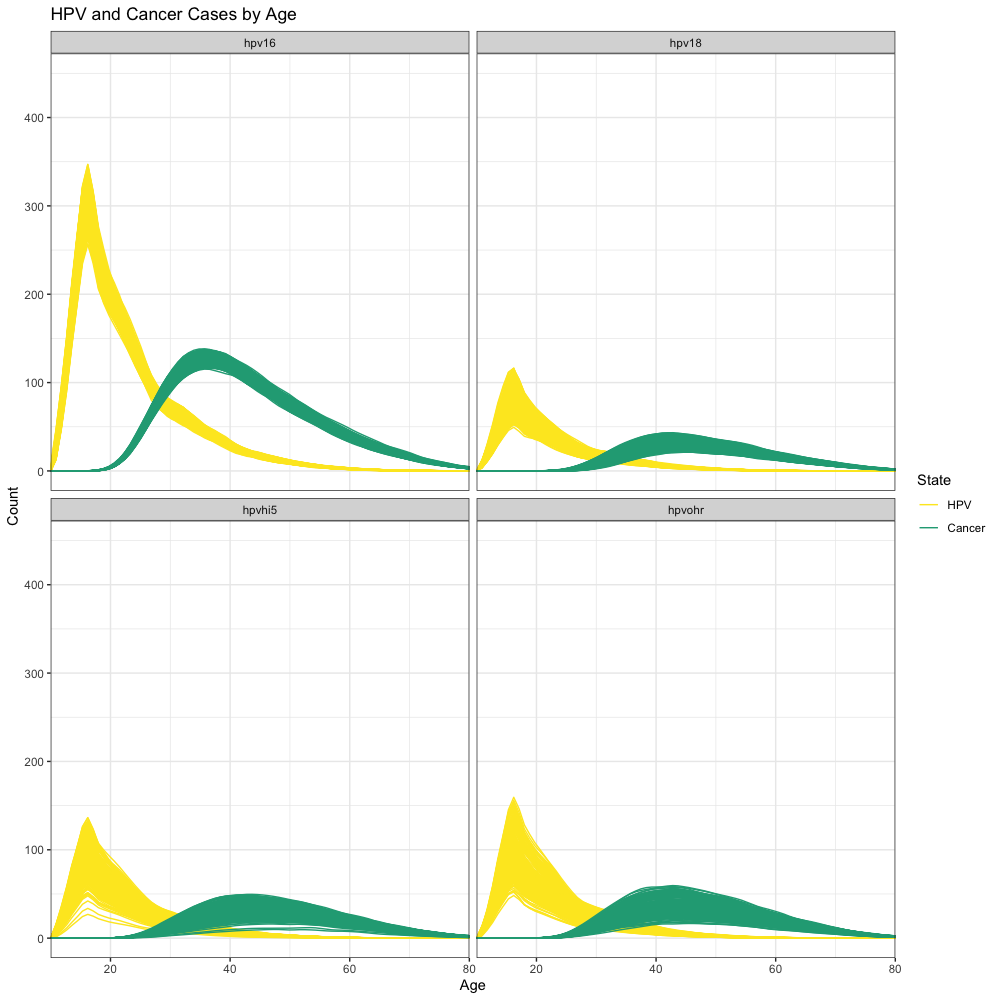}
    \caption{For each genotype class, the ages at which HPV is contracted and progresses to cancer.}
    \label{fig:hpvcancergeno}
\end{figure}

\begin{figure}
    \centering
    \includegraphics[width=\columnwidth]{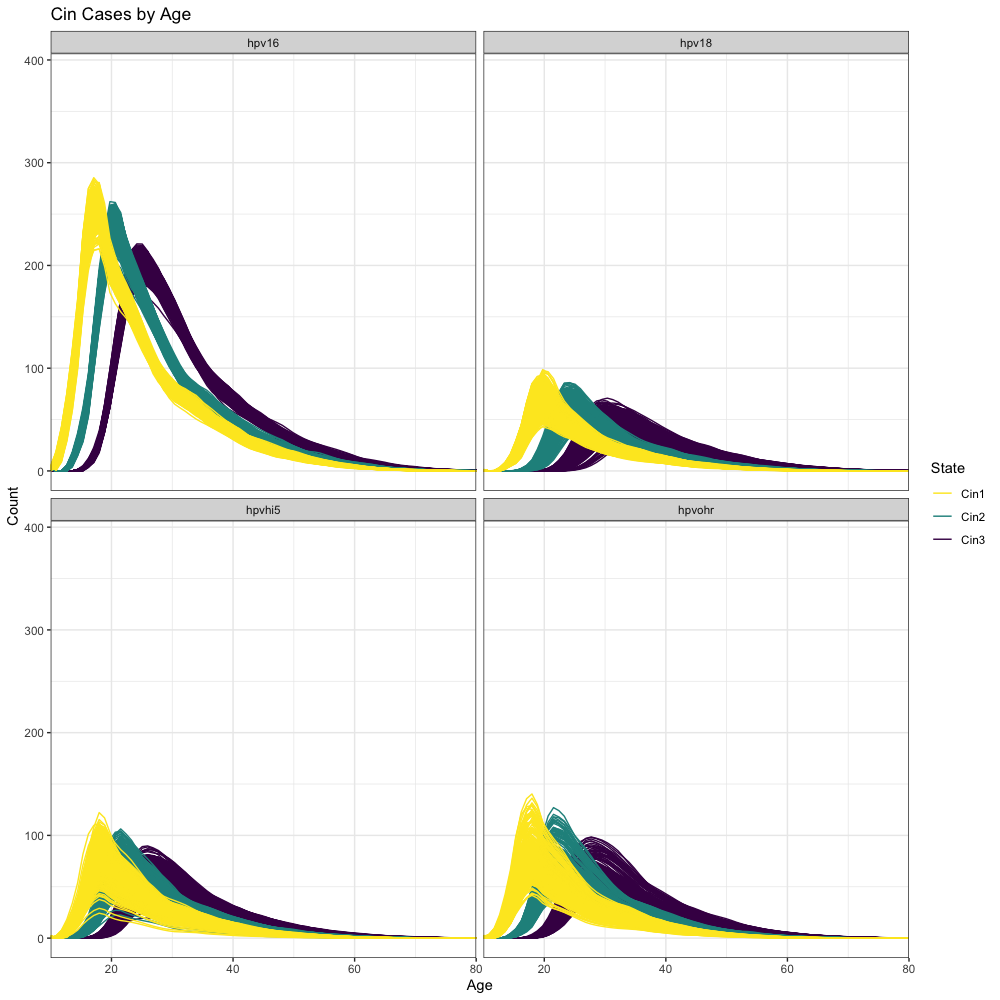}
    \caption{For each genotype class the ages of acquisition of intermediate CIN stages.}
    \label{fig:hpvcingeno}
\end{figure}

We first consider the acquisition and progression of HPV and its transformation to cancer. Due to the agent-based nature of \texttt{HPVsim}, we may interrogate the progression of each individual from the moment they contract HPV to their clearance of the disease or transformation to cancer, thus building up a demographic study of the ages at which HPV is contracted and the time taken to progress through each of the cervical intraepithelial neoplastic stages. This information could be of extreme value, potentially in conjunction with targeted longitudinal studies, to evaluate the optimal ages at which screening or vaccination strategies should be focused. The results are shown in Figures~\ref{fig:hpvcancergeno} and \ref{fig:hpvcingeno}.


We can see that the impact of each of the genotype classes on the progression of HPV to cancer are similar in shape, but there are some key differences between them. The spreads of possible stage acquisitions for the individual genotypes, HPV16 and HPV18, are much tighter than those for the $5$-genotype classes HPVhi5 and HPVohr. This is unlikely to be due to the fact that the $5$-genotype classes consist of multiple different strains of HPV, since \texttt{HPVsim} models those classes in the same fashion as it does the individual genotypes; moreover the flexibility in our targets used (and described in Supplementary Material C) for these genotype classes is no less restrictive than for the individual genotypes. We may conclude that, in order to obtain the observed data we have available to us, the individual genotype classes have a much more restricted possible progression path within the population.

We may also consider general behaviours between these genotypes, particularly in considering the most likely length of time spent in each state. While each HPV genotype reaches peak prevalence in the population at around the same time, at the ages of $\ageto{15}{17}$, there is a marked difference in the equivalent age for cancer. Table~\ref{tab:HPVprogress} shows the most common times spent in each state.

\begin{table}[!ht]
\centering
\begin{tabular}{r|c|c|c|c|c}
     \textbf{Genotype} & \textbf{Precin} & \textbf{CIN1} & \textbf{CIN2} & \textbf{CIN3} & \textbf{Total} \\
     \hline
     HPV16 & $1$ & $4$ & $5$ & $11$ & $21$ \\
     HPV18 & $4$ & $5$ & $8$ & $11$ & $28$ \\
     HPVhi5 & $2$ & $4$ & $6$ & $17$ & $29$ \\
     HPVohr & $2$ & $5$ & $6$ & $15$ & $28$ \\
\end{tabular}
\caption{The respective times (in years) spent in each disease stage from contracting HPV to full transformation to cancer -- here ``precin'' represents time spent with HPV without sufficiently many cells transformed to qualify as having cervical intraepithelial neoplasia.}
\label{tab:HPVprogress}
\end{table}

While there is little material difference between genotypes when considering time spent in a particular state, the cumulative effect of the more severe genotypes is significant. In particular, the modal age at which cancer is present for an individual infected with HPV16 is $36$, in comparison with $\ageto{43}{46}$ for all other genotypes. If we consider the `precin' and low-grade lesion (CIN1) states as being manageable (via pharmaceutical treatment or self-clearance), then this suggests that it is paramount that any screening procedures focus on the time spent in these stages: namely, ages $\ageto{15}{24}$. Once an individual has reached CIN2, it is unlikely that an individual will self-clear and prevent progression to later stages and, eventually, cancer. In these circumstances more involved, expensive, and painful treatments are required, making early detection paramount \cite{marth2017cervical}.

\begin{figure}[!ht]
\centering
\includegraphics[width=\columnwidth]{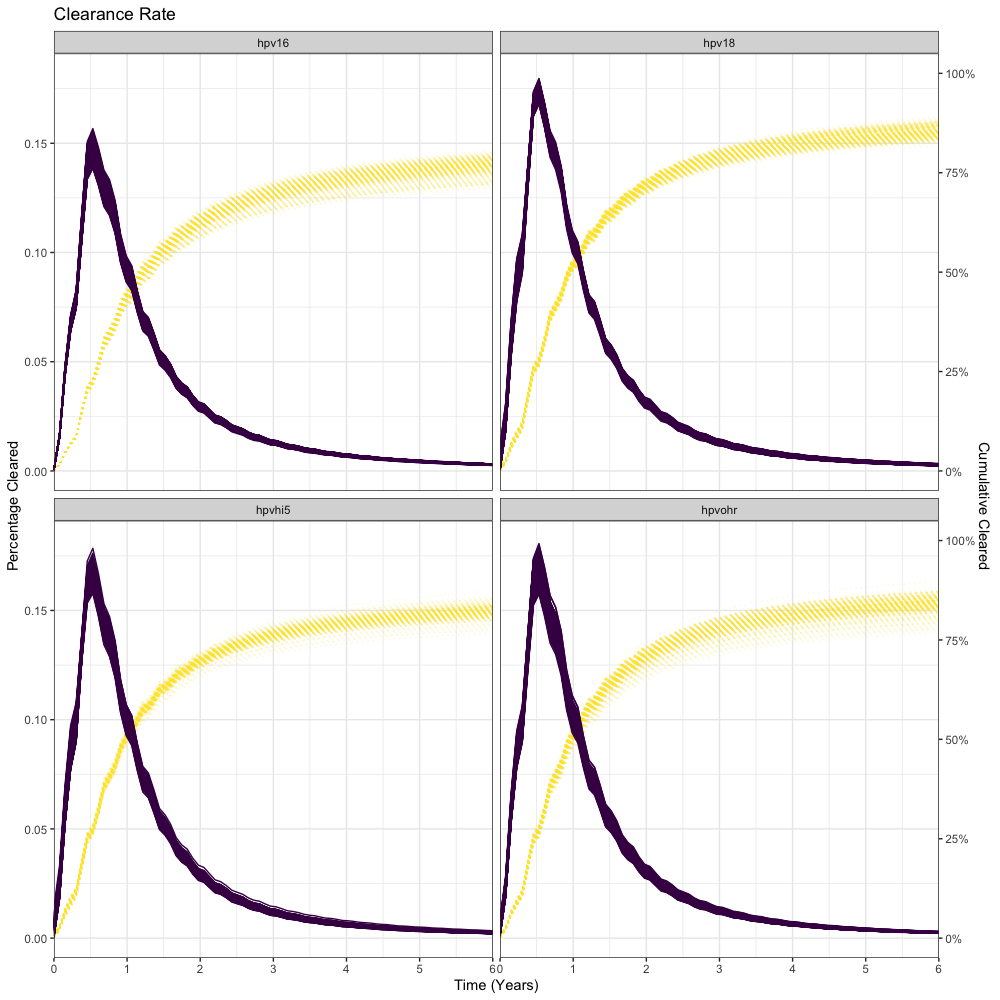}
\caption{Time taken to naturally clear HPV, by genotype. Solid lines are proportions of infected individuals clearing, by time taken to clear; dotted lines indicate cumulative proportions of infected individuals having cleared the disease by the given time.}
\label{fig:clearance}
\end{figure}

As mentioned, it is possible for individuals who contract HPV to clear the disease of their own accord. In this case, we may be interested in the time taken to clear the disease: in particular, whether the clearance time is lower than the time taken to reach the CIN2 stage. Figure~\ref{fig:clearance} shows the proportion of individuals that clear HPV in a certain number of years, collected in quarter-year intervals, as well as the cumulative proportion of infected individuals that clear the disease over time.

We may note that the peak of disease clearance is around $6$ months, common to all genotypes under consideration; after $1$ year, approximately $50\%$ of infected individuals have cleared their infection. This is in line with observational studies on self-clearance of HPV, which suggests a median clearance time of $13.5$ months and $52\%$ of cases cleared within this period \cite{adebamowo2018clearance}\footnote{This data is an appropriate comparator as it ignores any external contributions from HIV positivity or treatment pathways, in alignment with our modelling scenario.}. One interesting distinguishing feature is in the cumulative clearance rate from each genotype: we see that HPV16 is more likely to persist beyond $6$ years than the other genotypes. Coupled with the comparatively short gestation time seen in Table~\ref{tab:HPVprogress} we might consider this genotype to be the critical strain of interest, and a potential focus for any further study.

\begin{figure}[!ht]
\centering
\includegraphics[width=\columnwidth]{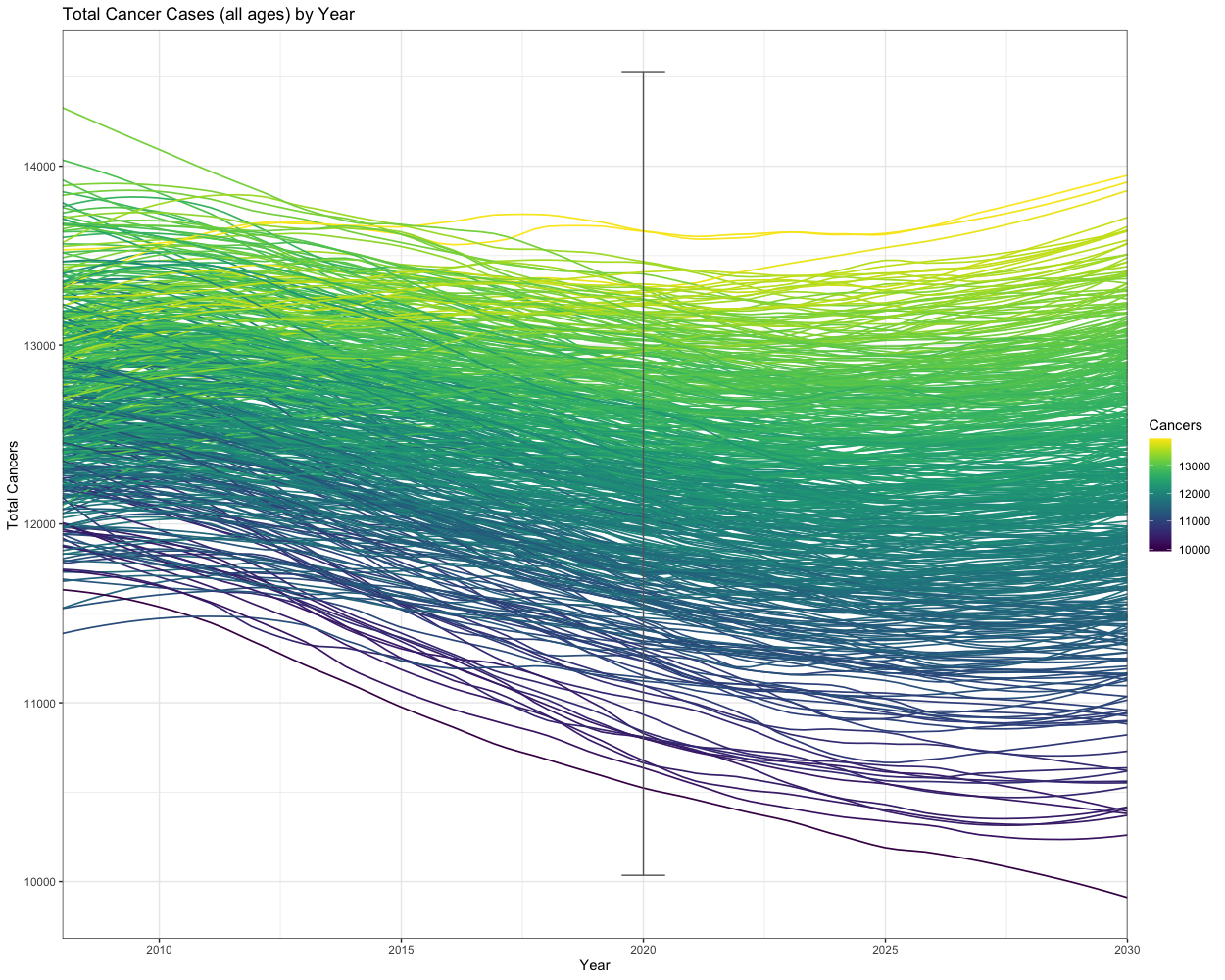}
\caption{Projected total cancer cases for the period $\ageto{2010}{2030}$. An error-bar at $t=2020$ represents the aggregated total number of cancers from observational data.}
\label{fig:cancerpreds}
\end{figure}

As a final example, we briefly consider the range of possibilities in the future. We do not make definitive statements save to highlight the breadth of possible outcomes given the observational data we have had access to. Figure~\ref{fig:cancerpreds} is derived from running \texttt{HPVsim} at each of the final proposal points across time, and considering total number of cancer cases at each year from $2010$ to $2030$.

We note that all points proposed, based as they were on age-aggregated targets, nevertheless also fall within the overall total number of cancers for $2020$, as demonstrated by the vertical error-bar. This needn't have been the case: for example, had some of the parameter combinations resulted in agreement with all observational data but always as an over-estimate of the means, then we may have seen trajectories above this bar (a point we return to shortly). We may see, nevertheless, that there remains a large spread on this higher-level statistic, and that spread propagates to any predictions we might make in the future. Averaging across predictions at each of $2020$ and $2030$, there is little difference ($12{,}300$ versus $12{,}275$), and the proportion of predictions that suggest a decrease in cancer cases over the ten year period is $\sim56\%$, giving no strong indication about future behaviour. The projected change is in the range $[-5.7\%, 4.5\%]$, corresponding to between $600$ fewer and $580$ more cancer cases in $2030$ compared to $2020$. This suggests a final range of total cancers in $2030$ of $[9{,}900, 14{,}000]$. The two extremal predictions are almost directly opposite, and if taken in isolation would result in entirely different conclusions, and yet they are of equal validity given the information available to us\footnote{For reference, the equivalent analysis on the optimisation runs mentioned in the preceding section suggests a projected change in the range $[-4\%, 1.7\%]$ corresponding to between $450$ fewer and $250$ more cancer cases in $2030$, with $30\%$ of proposed points suggesting an overall increase in numbers of cancer cases during this time.}. Were we to consider using \texttt{HPVsim} to make predictions of efficacy of interventions, the range of possible baseline outcomes must be taken into account in order to robustly predict the effect of any collection of vaccination and screening strategies, and certainly when considering cost-effectiveness of those strategies.

Given that the range that the predictions can span is almost certainly of critical importance for a robust analysis, we might consider further whether the points obtained through history matching represent the full range of behaviours. There may be a section of a boundary in the high-dimensional space which is not included in this ensemble of final runs which would give rise to a more extreme prediction -- since we are exploring a $33$~dimensional space we would not expect the ensemble of $500$ points here to explore all corners and boundaries of the non-implausible space and know that extreme behaviour is likely to fall on the boundary. We can, however, use these points as a training set for emulating the progression of total cancer cases over time and thence seek to find the projected maximum and minimum via the emulator, rather than the simulator.

\begin{figure}[!ht]
\centering
\includegraphics[width=\columnwidth]{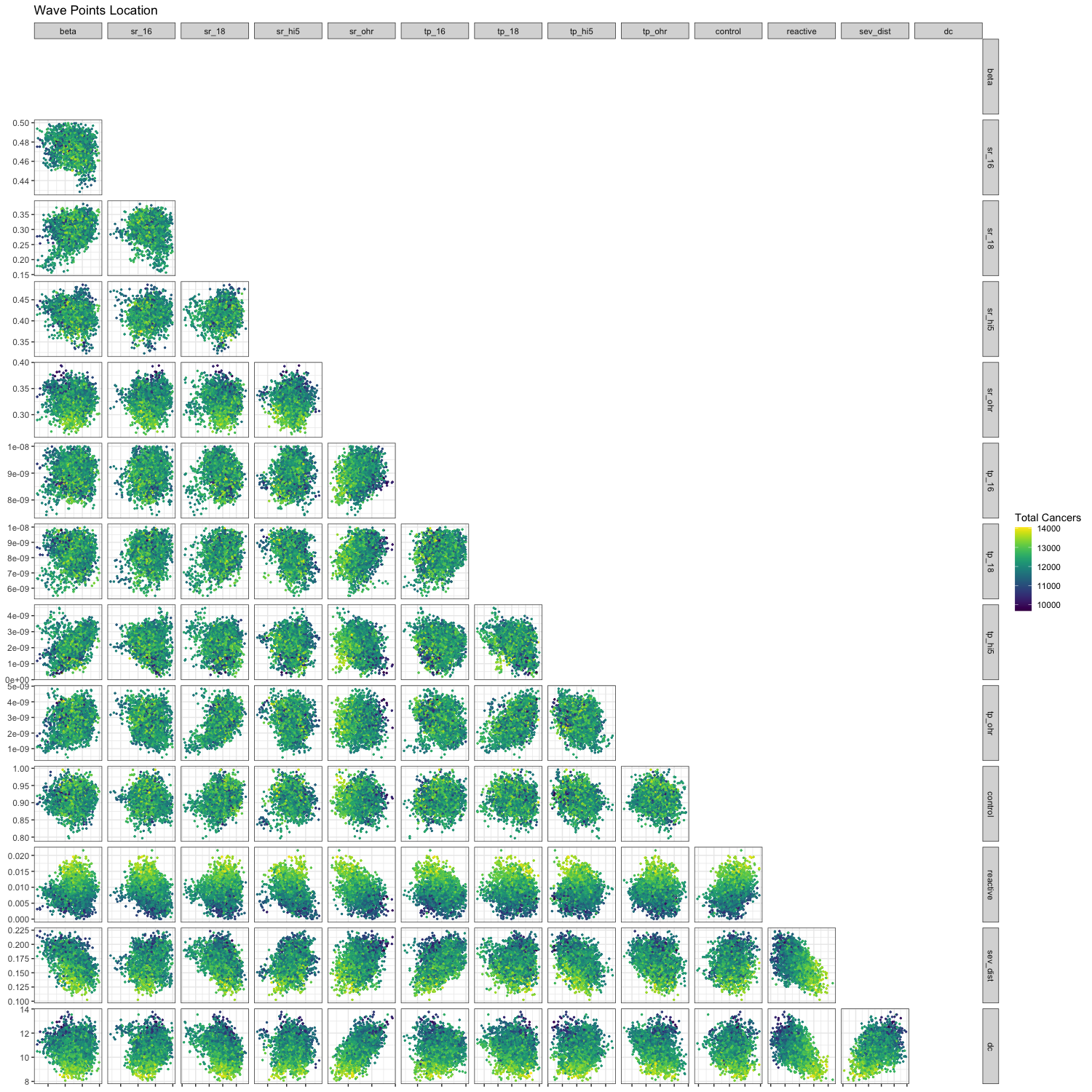}
\caption{Pairs plots of $5{,}000$ points from the final non-implausible region, coloured by projected number of cancers in $2030$. For clarity, only a subset of the $33$ parameters have been included as in Figure~\ref{fig:wavepoints}.}
\label{fig:cancerpoints}
\end{figure}

Figure~\ref{fig:cancerpoints} gives us an indication of the behaviour of the total number of cancer cases across the entirety of the non-implausible space. In order to aid the investigation of this, we have used our final wave of emulators to propose $5{,}000$ points from the non-implausible region. We then created an emulator for the simulator prediction of cancer cases in $2030$ using the $500$ runs above, and used this trained emulator to predict over the larger ensemble of parameters. In contrast with the equivalent task using \texttt{HPVsim}, the process of proposal, training, and prediction required less than an hour's computational time.

This gives us some degree of intuition into the drivers of high numbers of cancer cases in $2030$: in particular, high reactivation rate \texttt{reactive} has an obvious effect on increasing the numbers of cancers in the future as we would expect. There are also interesting features here which may bear consideration; in particular, the negative correlation between cancer cases in $2030$ and both duration of cancer (\texttt{dc}) and individual-level severity (\texttt{sev\_dist}). Such features may be related to behavioural changes in sexual contact when individuals suffer from cancer for a longer period of time, or where they are infected with more severe cases of HPV and thus progress to diagnosable cases more quickly. Fundamentally, however, we find that over this much larger space of acceptable parameter combinations the predictions of change in cancer cases from $2020$ to $2030$ is widened further, but not excessively so: according to our trained emulator, projected changes range from $-7.5\%$ to $+5.8\%$, corresponding to between $800$ fewer and $750$ more cases; the overall range of possible total cancers in $2030$ is $[9{,}600, 14{,}100]$. Fundamentally, this potential difference of $4{,}500$ cancer cases will have a huge effect on any decision making we would wish to perform and model using a simulator such as \texttt{HPVsim}. One may see that decisions made about clinical interventions or the focus of new screening programs would drastically change were we to consider only one of the extremes of predicted behaviour; the framework of emulation and history matching allows clinicians and economists to be aware of the complete space of possible effects any intervention may have, and make appropriately informed decisions as a result.

\section{Discussion}\label{sec:discuss}

Models such as \texttt{HPVsim} are invaluable in investigating complex infections at a national level, allowing for an exploration of the natural history of a disease that is not possible from observational studies. Any shortfalls in collected data that we do possess can, in many cases, be mitigated by inference from such a model; parameter combinations that provide matches to this observational data can be used in a variety of ways to analyze physical characteristics of the disease, national demographics, or consider the potential effects of interventions. However, the complexity inherent in using such a model for a thoughtful analysis of a disease can preclude a meaningful and thorough exploration of the parameter space, particularly when the parameter space in question is of high dimension.

We have argued here that emulation and history matching addresses the problems in matching a complex model to observational data efficiently. The application of this framework to \texttt{HPVsim}, facilitated by the R package \texttt{hmer}, allowed us to start with very uninformative parameter ranges (restricted only by physical impossibility, rather than expert judgement on their true values) and substantially reduce it to find a large collection of matches to observed data with comparatively few simulator evaluations: we required fewer than $200{,}000$ simulator evaluations in total to identify an acceptable subspace approximately $17$ orders of magnitude smaller than our original space. In the process of performing our waves of emulation, we obtained over $800$ parameter combinations whose simulator evaluations provided matches to our data, given the underlying uncertainties in observation, model, and stochasticity; having performed the $16$ waves of emulation we can generate many more such parameter combinations in a short space of time. In contrast, performing optimisation in the same setting required many more simulator evaluations to obtain a handful of final proposal points; these final parameter combinations were still outperformed by those proposed by history matching in almost all cases, both when visually analyzing these proposals and by the optimisation scheme's own goodness-of-fit measure. More importantly, the parameter sets proposed by history matching implicitly take into account the imperfections inherent in any simulation task, which is non-trivial (or in some cases, impossible) to include in other methodologies such as optimisation.

In the context of \texttt{HPVsim}, the large collection of parameter sets proposed from history matching comprise a representative collection of the complete parameter space that could give rise to our physical reality; this space can be used to make further inference about unobservable characteristics of the disease, make meaningful comparisons between different genotypes therein, and form a basis for robust prediction of future trends and response to intervention. We have identified a collection of factors that would help with assessing the implications of future behavioural changes while ensuring the validity of any inference drawn. In this work we have focused on the natural history of HPV in the community, particularly the acquisition, clearance, and progression of HPV to cancerous cells, but one could apply this process to any possible output of the simulator. We have also considered the output space of possible cancer cases in the future, in the absence of intervention, in order to highlight the range of different baseline scenarios that should be considered in an intervention scenario. Of most interest is the fact that our output space is split almost half-and-half as to whether we expect annual cancer cases to decrease or increase in $2030$ compared to our observations in $2020$; this is of key importance were we to consider applying interventions and analysing their effect.

The eventual uses of the calibrated model can be instructive in ways other than that described here. While we considered the dwelltime of genotypes as being unobservable in Section~\ref{sec:analysis}, one could see such quantities as in-principle observable given new studies of the disease. One might use the information gained here to identify observational quantities that would be most informative in further reducing the non-implausible region to aid further study design. This concept of using emulation to guide experimental design has been touched upon in other settings \cite{jackson2018design}, but would be of paramount importance here. Large-scale population studies of disease are time-consuming and expensive and, frequently, those quantities that are most accurately observed are not necessarily those that we would select for divining properties of the disease. By using this framework to inform data-gathering campaigns, we may be able to provide concrete statements about the most important information required to characterise the disease in a given country. We hope to explore this avenue of research in future works.

Such avenues of exploration and investigation notwithstanding, the work presented here provides a blueprint for robust examination of complex, stochastic, computer models and a means by which we can interrogate the final space of acceptable matches. With the increase in availability of computational resources, agent-based models such as \texttt{HPVsim} are common in modelling communities, and their use is likely to increase in the future. We hope that the techniques demonstrated in this work provide a framework for modelling communities to perform comprehensive analyses of their simulators, make confident statements about inferential quantities of interest, and provide robust predictions and recommendations for policy makers that can materially improve the landscape of global health.

\section*{Acknowledgements}
AI, DS, and NM would like to acknowledge the support and funding provided by the Wellcome Trust grant 218261/Z/19/Z; AI is supported by Durham University's Willmore Fellowship. RGW is funded by the Wellcome Trust (310728/Z/24/Z, 218261/Z/19/Z), NIH (1R01AI147321-01, G-202303-69963, R-202309-71190), EDTCP (RIA208D-2505B), UK MRC (CCF17-7779 via SET Bloomsbury), ESRC (ES/P008011/1), BMGF (INV-004737, INV-035506), and the WHO (2020/985800-0).

\bibliographystyle{model1-num-names}

\bibliography{references}

\clearpage

\onecolumn
\appendix
\section{Bayes Linear Emulation of Stochastic Models}\label{app:covem}
\subsection{Bayes Linear Emulation}
Here we provide details of the process behind emulation, particularly as applied to emulating the stochasticity of agent-based models. Further details may be found in \cite{vernon2022bayes, goldstein2007bayes}.

Consider a simulation of a real-world process $y$ which takes a set of input parameters, described as a vector $x$ of length $d$, and returns a set of $m$ outputs $\{f_i(x)\}_{i=1,\dots,m}$. A univariate emulator is a fast statistical approximation of one of the simulator outputs, built using a comparatively small collection of simulator runs, which provides an expected value for the simulator output at any unseen points $x$ in the parameter space along with a corresponding estimate of the uncertainty of the prediction, reflecting our beliefs about the simulator in question.

Concretely, we create a prior representation of a given simulator output $f_i(x)$ in emulator form as
\begin{equation}\label{eq:emulator}
g_i(x) = \sum_{j=1}^{p_i} \beta_{ij} h_{ij}(x_{A_i}) + u_i(x_{A_i}) + w_i(x).
\end{equation}
Here $x_{A_i}$, where $A_i\subseteq\{1,\dots,d\}$, are the set of `active variables' for output $f_i(x)$; that is, the components of the input vector $x$ that are most influential in determining the behaviour of $f_i(x)$. The $h_{ij}$ are $p_i$ known simple functions of the $x_{A_i}$, with $\beta_{ij}$ the corresponding coefficients -- together these two terms define a regression surface encoding the global response of $f_i(x)$ to the inputs. $u_i(x_{A_i})$ is a second-order weakly stationary process which captures residual variation in the active variables and can be seen as governing the local behaviour of the simulator output. While the regression functions $h_{ij}$ are considered known, we view their coefficients and the second-order process as being unknown and therefore treat them as random variables. We assume the following covariance structure for $u_i(x_{A_i})$:
$$\Cov[u_i(x_{A_i}), u_i(x^{\prime}_{A_i})] = (1-\delta_i)\sigma_i^2 r(x_{A_i}, x^{\prime}_{A_i}).$$
$r(x, x^{\prime})$ is a correlation function, suitably chosen depending on our beliefs about the output (common examples and their usage can be found in \cite{rasmussen2003gaussian}); our choice often depends on how `smooth' we expect the output response to be with respect to our input parameters and the extent to which neighbouring points in parameter space influence the point $x$ in question. Our overall variance $\sigma^2_i$ represents our prior uncertainty about our predictions, and $\delta_i\in[0,1]$ accounts for the residual effect of the inactive variables on the simulator output. This is accounted for in the `nugget' term $w_i(x)$, which has correlation structure
$$\Cov[w_i(x), w_i(x^{\prime})] = \sigma_i^2 \delta_i I_{x=x^{\prime}}$$
with $I_{x=x^{\prime}}$ is an indicator function taking the value $1$ if $x=x^{\prime}$ and $0$ otherwise.

If we further assume that the regression coefficients $\beta_{ij}$, second-order process $u_i$, and nugget term $w_i$ are mutually uncorrelated, that $\Exp[u_i(x)]=\Exp[w_i(x)]=0$ for all $x$, and that $\Var[\beta_{ij}]=0$ \cite{goldstein2013assessing}, then we see that our emulator prediction at a point $x$ is determined by the regression surface, with variance in that prediction given by $\sigma^2$:
$$
    \Exp[g_i(x)] = \sum_{j=1}^{p_i}\Exp[\beta_{ij}]h_{ij}(x_{A_i}), \hspace{5mm}\Var[g_i(x)] = (1-\delta_i)\sigma_i^2 + \delta_i \sigma_i^2 = \sigma_i^2.
$$
The choice of a zero-variance regression surface decouples the global behaviour of the simulator from the weakly-stationary process, allowing the latter to focus on representing the variation of the residuals, providing a clear separation between the global and local behaviour and allowing for a convenient representation of our beliefs about the output via the regression surface and our uncertainty about those beliefs.

Before we can update the emulated structure with respect to data, we must complete the \textit{a priori} specifiction for the quantities $\beta_{ij}$, $u_i(x_{A_i})$, and $w_i(x)$. This can be done in a variety of ways: for instance, if one is willing and able to specify full distributions for these quantities, we could then use maximum likelihood (ML) or maximum \textit{a posteriori} (MAP) estimates to determine plug-in values for their hyperparameters \cite{andrianakis2012effect}, or use cross-validation \cite{maatouk2015cross}. However, we may not be able (or willing) to make full distributional specifications for these quantities, whether due to a lack of prior knowledge required for such a specification or a lack of faith in such statements. It is rare, however, that we are similarly hampered in making statements about the second-order structure of the system -- expectations and (co)variances -- particularly in light of the emulator structure described above. The Bayes linear framework requires only these quantities, and we proceed accordingly. Explicitly, then, we focus on specifying the quantities $\Exp[\beta_{ij}]$, $\sigma_i$, $\delta_i$, and any hyperparameters that govern the behaviour of the correlation function $r(x, x^{\prime})$. These quantities can be determined via a full \textit{a priori} specification or by using pragmatic plug-in estimates \cite{santner2003design, kennedy2001bayesian, rasmussen2003gaussian}.

We may now use data obtained from simulator runs to update our beliefs in light of data. Suppose that we have data obtained from running the simulator at a series of points $(x^{(1)}, x^{(2)}, \dots, x^{(n)})$, resulting in a collection of outputs\footnote{For a stochastic simulator, we would obtain multiple output values for a given input parameter $x$; we will discuss this point momentarily.}
$$D_i = \left(f_i(x^{(1)}), \dots, f_i(x^{(n)})\right).$$
If we had a full distributional statement about our emulator quantities, then we would produce a posterior estimate using Bayes' Theorem; the equivalent in our case is given by the Bayes linear update formulae \cite{goldstein2007bayes}. The Bayes-adjusted emulator prediction for $g_i(x)$ at new points $x$, $x^{\prime}$ is given by
\begin{align}
    \Exp_{D_i}[g_i(x)] =& \Exp[g_i(x)] + \Cov[g_i(x), D_i]\Var[D_i]^{-1}(D_i - \Exp[D_i]),\label{eq:blupexp} \\
    \Cov_{D_i}[g_i(x), g_i(x^{\prime})] =& \Cov[g_i(x), g_i(x^{\prime})]- \Cov[g_i(x), D_i]\Var[D_i]^{-1}\Cov[D_i, g_i(x^{\prime})]\label{eq:blupcov}
\end{align}
where here $\Exp_{D}[g_i(x)]$ is the expectation or prediction for the output in question, updated with respect to the data $D$, $\Cov_D[g_i(x), g_i(x^\prime)]$ is the adjusted covariance, and $\Var_D[g(x)]=\Cov_D[g_i(x), g_i(x)]$ is the adjusted variance corresponding to this prediction.

We therefore obtain a prediction and corresponding uncertainty for the value of $f_i(x)$ at a point $x$ whose simulator output has not been evaluated. The uncertainty depends on our prior beliefs and the proximity of data obtained to our unseen point; indeed one can show that for a deterministic simulator, the posterior prediction at a `training' point $x^{(k)}$ is identical to the simulator output $f(x^{(k)})$ with vanishing posterior variance in the absence of a nugget term. Crucially, the quantities \eqref{eq:blupexp} and \eqref{eq:blupcov} are extremely fast to evaluate, so the emulators allow us to perform extensive exploration of the simulator's behaviour over the input space.

As a final note, were we to have assumed normal and Gaussian process priors for $\beta$ and $u(x)$, respectively, then the approach here bears close similarity to Gaussian process emulation \cite{conti2009gaussian}. However, Gaussian process emulation requires invoking additional distributional assumptions that may be difficult to justify, force stricter and more complex diagnostics to be satisfied, and materially affect our final inference. The Bayes linear approach allows us to circumvent these concerns and provide meaningful statements about the behaviour of our simulator without invoking potentially unjustified assumptions about the underlying distribution.

\subsection{Covariance Emulation}
As alluded to, the process of adjusting our emulators above presupposes that we obtain a single output value from application of the simulator to each parameter set $x^{(i)}$. For a deterministic model, this is all that is required to understand the output surface at the points of interest; for a stochastic model, however, repeated evaluations of the simulator at the same parameter set will not be identical. We therefore propose a modification to the framework presented above.

Suppose that we have two simulator outputs $f_a(x)$ and $f_b(x)$ for a parameter set $x$. Performing $n$ evaluations of the simulator at $x$ gives rise to realisations $(\mb{f}{1}_a(x), \mb{f}{2}_a(x), \dots \mb{f}{n}_a(x))$, and similarly for output $b$. For each output, we assume that realisations are second-order exchangeable, so that
$$\mb{f}{k}_a(x) = \mathcal{M}(f_a(x)) + \mathcal{R}_k(f_a(x)),\hspace{2mm}k=1,2,\dots,n.$$
$\mathcal{M}(f_a(x))$ can be thought of as the true mean of the output $f_a(x)$ that would be observed were we to perform infinitely many evaluations at the point $x$, while $\mathcal{R}_k(f_a(x))$ represents the residual variability due to the $k^{th}$ realisation from this mean. We may extend the concept of second-order exchangeability to the residuals; for outputs $a$ and $b$ we define
$$\mathcal{R}_k(f_a(x))\mathcal{R}_k(f_b(x)) \equiv Q^{(k)}_{ab}(x) = \mathcal{M}(Q_{ab}(x)) + \mathcal{R}_k(Q_{ab}(x)),$$
where now $Q_{ab}$ represents the true covariance between outputs $a$ and $b$. Note that, if $a=b$ then $Q^{(k)}_{ab}(x) = \left(\mathcal{R}_k(f_a(x))\right)^2$ and the specifications are identical to those proposed in \cite{vernon2022bayes}.

It is not feasible to observe the true mean from the simulator, nor the true covariance matrix: for finitely many realisations at the point $x$ we may obtain the sample means and covariances. Let the sample covariance obtained from $n$ realisations be denoted $q_{ab,n}$: then computation similar to that given in \cite{goldstein2007bayes} yields the following relation:

$$
    q_{ab,n}(x) = \frac{1}{n}\sum_{k=1}^n \left(\mathcal{M}(Q_{ab}(x)) + \mathcal{R}_k(Q_{ab}(x))\right) - \frac{1}{n(n-1)}\sum_{k\neq l}\mathcal{R}_k(f_a(x))\mathcal{R}_l(f_b(x)) \equiv \mathcal{M}(Q_{ab}(x)) + T_{ab}(x).
$$
Hence the sample covariance is related to the true covariance $\mathcal{M}(Q_{ab}(x))$ via the additional term $T_{ab}(x)$, where explicitly
$$T_{ab}(x) = \frac{1}{n}\sum_{k=1}^n \mathcal{R}_k(Q_{ab}(x)) - \frac{1}{n(n-1)}\sum_{k\neq l} \mathcal{R}_k(f_a(x))\mathcal{R}_k(f_b(x))$$
can be thought of as a correction term that reconciles the two quantities. Under the assumptions of second-order exchangeability; namely that residuals are mutually uncorrelated and uncorrelated with the true mean and covariance, we find that the additional variability due to finite sample size is
$$
    V^{ab}_T(x) \equiv \Var[T_{ab}(x)] = \frac{1}{n} C_{ab, R(C)}(x) + \frac{1}{n(n-1)}\left(\Cov[\mathcal{M}(Q_{aa}(x)), \mathcal{M}(Q_{bb}(x))]\right. \left.+C_{aa}(x)C_{bb}(x) + C_{ab,M}(x) + C_{ab}(x)^2\right),
$$
where for ease of notation we have defined
\begin{align*}
    C_{ab}(x) &= \Exp[\mathcal{M}(Q_{ab}(x)], \\
    C_{ab,M}(x) &= \Var[\mathcal{M}(Q_{ab}(x)], \\
    C_{ab, R(C)}(x) &= \Var[\mathcal{R}_k(Q_{ab}(x))].
\end{align*}
Note that, for increasing $n$, the contribution of the correction term decreases; in the limit as $n\to\infty$, the correction term vanishes as we would expect. Furthermore, the nature of the correction term is dominated by the residuals on the covariance structure, $\mathcal{R}_k(Q_{ab}(x))$, not the residuals on the mean surface; though both play a part for modest $n$.

Given determinations about the relevant prior quantities, and with the corresponding emulators for the mean and variance constructed using those priors, the update formulae presented in \eqref{eq:blupexp} and \eqref{eq:blupcov} follow straightforwardly, save for one key aspect. For simplicity, consider only two simulator outputs and suppose that we have data points $(x^{(1)}, \dots, x^{(m)})$ upon each of which we perform $n_l$ repetitions, $l=1,\dots,m$ giving rise to sample means $D_{u}=(\mu_u^{(1)}, \dots, \mu_u^{(m)})$ and sample (co)variances $S_{uv}=(q_{n_1}^{uv}(x_1), \dots, q_{n_m}^{uv}(x_m))$ for $u,v\in\{a,b\}$. Let $V_{M,i}^{uv} = \Var[\mathcal{M}(Q_{uv}(x^{(i)}))]$ and $C_{M,ij}^{uv} = \Cov[\mathcal{M}(Q_{uv}(x^{(i)})), \mathcal{M}(Q_{uv}(x^{(j)}))]$, then for each element of the covariance matrix we seek to emulate we have

$$
\Var[S_{uv}] = 
\begin{pmatrix}
V_{M,1}^{uv} + V_{T,1}^{uv} & C_{M,12}^{uv} & \dots & C_{M,1m}^{uv} \\
C_{M,21}^{uv} & V_{M,2}^{uv} + V_{T,2}^{uv} & \dots & C_{M,2m}^{uv} \\
\vdots & \vdots & \ddots & \vdots \\
C_{M,1m}^{uv} & C_{M,2m}^{uv} & \cdots & V_{M,m}^{uv} + V_{T,m}^{uv}
\end{pmatrix}.
$$

With this information in hand, we may perform the Bayes linear update \eqref{eq:blupexp} to obtain a posterior prediction for the elements of the covariance matrix. Once obtained, the mean data matrix $D$ obtains a similar corrective term based on this posterior prediction: let $V_u^*(x)$ be the adjusted prediction of variance for output $u$ at point $x$. Then
$$\Var[D_u] \to \Var[D_u] + \text{diag}\left(\frac{1}{n_1} V_u^*(x_1), \dots, \frac{1}{n_m} V_u^*(x_m)\right),$$
where $\text{diag}(\cdot)$ represents the diagonal matrix whose diagonal entries are the argument $\cdot$. The process of updating the mean predictions is unchanged save for this adjustment to the data variance matrix, and we may use the resulting emulators in the normal fashion.

There are two additional considerations when applying covariance emulation to a simulator. Firstly, we must determine second-order specifications for both the mean and variance emulator: this incorporates not just first- and second-order quantities, but also fourth-order quantities since for example $C_{ab,R(C)}(x)$ represents the variance of the residuals of a covariance. These higher-order quantities are less intuitive than those required to specify an emulator for the simulator output itself; while some can be determined in much the same way as for a mean emulator, given data, some cannot. A number of arguments about how to determine these quantities can be found in \cite{goldstein2007bayes}. Secondly, our implausibility measure is naturally modified by this change due to its dependence on our variance: in the multivariate form presented in \eqref{eq:multiimp} we obtain an extra term arising from covariance matrix $\Var[\mathcal{R}_k(f_a(x))]$.

Finally, one may note that this is not a true multivariate emulation; we are instead individually emulating the elements of a covariance matrix. For some applications, this could result in predicted covariance matrices that are not semi-positive definite, and we must be careful in such situations\footnote{Indeed, even if we were to simply emulate variances and ignore the off-diagonal terms of the covariance matrix, we may still predict negative variance at certain points in parameter space.}. One could treat such ill-posed predictions as diagnostic warnings on those parts of parameter space where they occur, effectively refusing to make statements about those parts of parameter space in the given wave; employ techniques to recast an ill-posed matrix as a valid covariance matrix, for example, performing an eigendecomposition and replacing any negative eigenvalues with $0$; or in some cases a transformation to log-covariance can be appropriate\footnote{Though one must be careful about the connection between the uncertainty in the prediction of the mean of the simulator output and the prediction of the mean of the log-variance.}. For systems with large numbers of outputs, this may also result in a prohibitively large number of emulators, since a system with $n$ outputs would result in the construction of $n + \frac{1}{2}n(n+1)$ emulators. Depending on the structure of the model in question, one might choose to focus on a subset of output covariances: for example, if we have good reason to believe that two outputs are (close to) uncorrelated, then no additional information would be gained from constructing a covariance emulator for that combination.

Intricacies notwithstanding, the framework presented above allows us to construct a natural hierarchy capable of encapsulating the output response and stochasticity over the parameter space, feeding informed statements about heteroskedasticity in the space of interest into posterior predictions of uncertainty on the outputs of interest.

\section{The History Matching Framework}\label{app:hm}
The emulators defined in Supplementary Material A have the benefit of being extremely fast to evaluate in comparison with the simulator they represent. The drawback of such an approach is that, by definition, the emulators are statistical \emph{approximations} of the simulator output; while they naturally provide a quantification of the uncertainty about any prediction made, this remains a source of uncertainty. However, this is by no means the only source of uncertainty in our calibration problem; a common aim when using simulators is to answer the following question, which implicitly describes those external sources of uncertainty and which we will further unpack shortly.

\emph{Given observed data corresponding to a simulator output, what combinations of input parameters could give rise to simulator output consistent with this observation?}
 
To answer this question, we apply the history matching approach \cite{craig1997pressure}, which aims to find the space of acceptable matches via complementarity. For further discussion of this choice and a comparison with other approaches, see \cite{bower2010galaxy, mckinley2018approximate, vernon2018bayesian}.

We must first create a link between an observation of a real-world process and our emulator. Let us denote one aspect of the real-world process, represented by simulator output $f_i(x)$, by $y_i$. Observations of $y_i$ are seldom made perfectly; for example, in epidemiological models it is common to expect that case numbers for a disease are subject to under-reporting or over-dispersion. We therefore link the observation $z_i$ of the process to the process itself via
$$z_i = y_i + e_i,$$
with $e_i$ some random quantity representing the observational error structure. Similarly, we would not expect the simulator to be a perfect representation of the physical reality it models: we link the two via
$$y_i = f_i(x) + \epsilon_i(x),$$
where $\epsilon_i(x)$ represents the (in general, $x$-dependent) `model discrepancy' for output $i$ at a parameter set $x$ which seeks to describe the deficiency between the simulator and reality \cite{goldstein2013assessing, brynjarsdottir2014learning, bower2010galaxy}. We also have a natural means by which to link the simulator output to the emulators as described in the preceding section; we therefore have a chain of statements that connect the observational data available to us to the emulator predictions.

This uncertainty structure allows us to approach the problem of finding acceptable parameter sets in a markedly different way to methods such as optimisation. Rather than seeking points in parameter space whose simulated outputs are likely to be good matches to the observational data (via, for example, maximising a goodness-of-fit measure), we instead focus on removing parts of parameter space that are highly unlikely to give rise to a good match. By `highly unlikely' here we mean that even accounting for all uncertainties in the system (emulator, observational, and model uncertainty), our prediction has a negligible probability of giving rise to our observation. We can therefore systematically and iteratively remove the unsuitable parts of parameter space, improve the emulators' predictions over the remaining space, and repeat the process until we arrive at the full parameter space of interest.

Concretely, we now define an \emph{implausibility} measure \cite{vernon2014galaxy} for observations $z = (z_1, \dots, z_m)$ corresponding to outputs $f_1(x), \dots, f_m(x)$. The implausibility for a single output $f_i(x)$ can be written as

\begin{equation}\label{eq:univimp}
I^2_i(x) = \frac{(\Exp_D[f_i(x)]-z_i)^2}{\Var_D[f_i(x)] + \Var[e_i] + \Var[\epsilon_i(x)] + \Exp_S[g_v(x)]},
\end{equation}
where $\Exp_D[f_i(x)]$ and $\Var_D[f_i(x)]$ are the prediction and variance for the mean response, and $\Exp_S[g_v(x)]$ symbolically represents the prediction of the stochasticity, relative to sample data $S$. If $I_i(x)$ is large, then we may think it unlikely that we would obtain an acceptable match to observed data were we to run the simulator at the point $x$; such a point is termed \emph{implausible}. Conversely, if $I(x)$ is small then we are unable to rule out the possibility that $x$ would give rise to a good match to observational data, given the information available to us; $x$ is correspondingly deemed \emph{non-implausible}. Note that the implausibility $I(x)$ can be small for two reasons: either the prediction $\Exp_D[f_i(x)]$ is close to the observation $z$, suggesting a good match to data; or else the uncertainties (particularly $\Var_D[f_i(x)]$) are large, suggesting that we currently do not have enough information to rule out a good match to data and that further investigation at this point would be useful. The implausibility measure therefore acts as a measure of both exploration and exploitation, indicating regions of parameter space whose consideration would improve our understanding of the simulator. A point is deemed implausible only if an simulator evaluation at this point has low probability of matching to observational data and inspection of the neighbourhood of the point would provide no additional insight.

In the presence of multiple outputs, a natural approach is to demand that a point be non-implausible with respect to all observations: this gives rise to a maximum implausibility measure
\begin{equation}\label{eq:maximp}
    I_M(x) = \max_i\{I_i(x)\}_{i=1,\dots,m}
\end{equation}
from which other, less restrictive, measures such as second-maximum $I_{2M}(x)$ and $n$th maximum implausibilities follow. We may combine these measures at will; for example, allowing at most one output to be slightly further from the data, but all to be within a certain closeness: we would therefore wish to satisfy a constraint $I_M(x) \le c_1$ on the maximum implausibility and an additional constraint $I_{2M}(x) \le c_2$ on the second maximum, with $c_1 > c_2$, for example.

We may also combine the multiple emulator implausibilities more directly, via a multivariate measure:
\begin{equation}\label{eq:multiimp}
    I^2(x) = (\Exp_D[f(x)] - z)^\top \left(\Var_D[f(x)] + \Var[e_i] + \Var[\epsilon_i(x)] + \Exp_S[f_v(x)]\right)^{-1} (\Exp_D[f(x)]-z) \\
\end{equation}
Here, the quantity $\Exp_D[f(x)]$ is taken to be the vector of predictions formed from the $m$ trained emulators of the outputs at a point $x$, $\Var_D[f(x)]$ the corresponding covariance matrix, and $\Exp_S[f_v(x)]$ the covariance matrix of the stochasticity, as predicted from the (trained) covariance emulators. This can be extremely useful when we have some understanding of the correlations between outputs, since this measure will constrain outputs according to their relations as well as their individual predictions.

The history matching approach proceeds in iterations. We refer to these iterations as `waves': at each wave $k$, a set of emulators are constructed for a collection of outputs $O_k$ based on a representative sample of points and their simulator evaluations from wave $k-1$. These trained emulators are used to assess implausibility over the space, $\mathcal{X}_k$, that remained at wave $k-1$, discarding those regions now deemed implausible to produce a representative sample of a smaller parameter space $\mathcal{X}_{k+1}$. These points, and their simulator evaluations, are used to inform emulators at wave $k+1$ and so on. The algorithm is shown in schematic form in Algorithm~\ref{alg:hm}.

\begin{algorithm}
    \caption{The History Matching Algorithm}\label{alg:hm}
    \begin{algorithmic}
    \Require Initial parameter region $\mathcal{X}_0$
    \State $k\gets1$
    \While{$\mathcal{X}_{k-1}\neq\emptyset$ and no other stopping condition is satisfied}
    \State Generate an appropriate design $\{x_i\}_{i=1,\dots,n}$ of $n$ points over $\mathcal{X}_{k-1}$
    \State Identify a collection of informative outputs, $O_k$
    \State Obtain simulator evaluations for outputs $O_k$ for each point $x_i$
    \State Construct emulators defined only over $\mathcal{X}_{k-1}$ for the collection $O_k$, trained using the simulator evaluations
    \State Calculate implausibility across the entirety of $\mathcal{X}_{k-1}$, identifying the new non-implausible region $\mathcal{X}_k$
    \State $k \gets k+1$
    \EndWhile
    \State If $\mathcal{X}_k\neq \emptyset$, generate a large number of acceptable runs from $\mathcal{X}_k$, sampled appropriately.
    \end{algorithmic}
\end{algorithm}

The stopping condition mentioned in Algorithm~\ref{alg:hm} depends on the ultimate aim of our model analysis; a point mentioned in the context of \texttt{HPVsim} in Section~\ref{subsubsec:hmdesc}. We pause here to highlight one potential stopping condition which is unique to the history matching framework: the possibility of stopping because the current non-implausible space is empty, $\mathcal{X}_k = \emptyset$ for some $k$. Unlike in methods such as optimisation and ABC, which return the `best' available point or posterior distribution, history matching is not similarly constrained and we may rule out the entire space as implausible. Such a situation can be extremely informative, representing a fundamental divergence between our observations, model, and reality, and force us to consider our error structure and model behaviour relative to the problem we wish to address.

The question of how to sample from a geometrically complex non-implausible region is an area of research in its own right; here we have followed the procedure detailed in \cite{iskauskas2022emulation} which we summarise here.
\begin{enumerate}
    \item Generate a large Latin hypercube and reject those points whose implausibility exceeds the cutoff;
    \item Use the points from Step 1 to select pairs of non-implausible points and draw rays connecting them. Add points that lay on these rays and on the boundary of the non-implausible region to the collection of points already obtained;
    \item Use the non-implausible set as the basis for a mixture distribution of uniform ellipsoids, performing importance sampling using this distribution;
    \item Thin the final collection of non-implausible points using a maximin argument, to produce a set of space-filling parameter combinations.
\end{enumerate}
This ensures, for most applications, that we obtain a collection of parameter sets that span the non-implausible region, and it proved to be adequate for point proposal in this application.

Just as we may choose outputs of interest, $O_k$ at each wave depending on the stability of the parameter space $\mathcal{X}_k$, so too may we modify our methodology for emulation. At early waves, we may deem the output covariance structure to be too volatile to provide any meaningful insight due to extreme behaviour in parts of parameter space we have yet to rule out; nothing precludes us from simply focusing on the emulation of means (with a suitably conservative estimate of stochasticity included as model discrepancy) until we have reduced the space to a more stable region of interest.

There is one final consideration that must be made in light of our modelling problem, which we briefly mention here. In the formulation for stochastic models and emulation thereof, we implicitly assume that we may match to the `true' means, $\mathcal{M}(f_i(x))$, of the system given sample data $\bar{f}_i(x)$ from a finite collection of realisations. The additional term in the denominator of the implausibility \eqref{eq:univimp} and inverse matrix of \eqref{eq:multiimp} accounts for the potential disconnect between the two quantities; this is only relevant if we believe that the true mean is of interest. As touched upon earlier, one may instead consider matching to individual realisations: for this we would instead wish to more carefully consider the covariance structure between outputs to adequately constrain the non-implausible space. We reserve this avenue of exploration for further work.

\section{\texttt{HPVsim} Parameters and Outputs}\label{app:details}
\subsection{Input Parameters}\label{app:detailsin}
$33$ parameters were chosen to be varied, and the initial allowed ranges determined via expert consideration of the physical system. Where a parameter name is of the form \texttt{name\_xx}, the \texttt{xx} corresponds to one of the four genotypes under consideration: $16$, $18$, $\text{hi}5$, or \text{ohr}.

\begin{center}
\begin{tabular}{|m{10em}|m{15em}|m{6em}|}
\hline
\textbf{Name} & \textbf{Description} & \textbf{Range} \\
\hline\hline
\texttt{beta} & Transmission probability & $[0.02, 0.25]$ \\
\hline
\texttt{de\_xx} & Mean (in years) of lognormal distribution of duration of infection prior to clearance, control, or transformation for genotype xx & $[3, 10]$ \\
\hline
\texttt{de\_p2\_xx} & Variance of lognormal distribution for episomal infection for genotype \texttt{xx} & $[5,15]$ \\
\hline
\texttt{sr\_xx} & Growth rate parameter mapping episomal duration to severity for genotype xx & $[0.1, 0.5]$ \\
\hline
\texttt{dp\_xx} & Mean (in years) of folded normal for precin duration of HPV for genotype \texttt{xx} & $[0.25, 4]$ \\
\hline
\texttt{tp\_xx} & Per-cell annual probability of transformation for genotype xx & $[10^{-11}, 10^{-8}]$ \\
\hline
\texttt{rb\_xx} & Relative genotype transmissibility to the baseline figure for genotype \texttt{xx} & $[0.7, 1.3]$ \\
\hline
\texttt{debut\_female} & Mean age of female sexual debut & $[12, 19]$ \\
\hline
\texttt{control} & Probability that hpv infection will be cleared by the host & $[0, 1]$ \\
\hline
\texttt{reactive} & Probability of reactivation of latent HPV infection & $[0, 0.15]$ \\
\hline
\texttt{sev\_dist} & Mean of folded normal for individual level severity scale factors & $[0.1, 2]$ \\
\hline
\texttt{pm} & Mean number of concurrent marital partners & $[0.001, 0.15]$ \\
\hline
\texttt{pc} & Mean number of concurrent casual partners & $[0.1, 0.6]$ \\ 
\hline
\texttt{dm} & Mean duration of marital partnership & $[8, 25]$ \\
\hline
\texttt{dc} & Mean duration of cancer & $[6, 15]$ \\
\hline
\end{tabular}
\end{center}

\subsection{Outputs and Uncertainties}\label{app:detailsout}
$22$ outputs were identified: of these, $14$ corresponded to age-stratified incidence of cancer in the year $2020$, and the remaining $8$ were proportions of the different HPV genotypes presenting in patients with cervical cancer and in those with high-grade lesions (\texttt{cin3}), measured in $2015$. Observation error is encoded via the intervals of the output values. Here, model discrepancy is entered as $\sqrt{\Var[\epsilon]}$ for each output.

\begin{center}
\begin{tabular}{|m{15em}|m{9em}|m{6em}|}
\hline
\textbf{Output name} & \textbf{Target} & \textbf{Model Disc.} \\
\hline
\texttt{cancer202015.0} & $[1, 38]$ & $0.812$ \\
\texttt{cancer202020.0} & $[183, 401]$ & $9.847$ \\
\texttt{cancer202025.0} & $[713, 1032]$ & $28.151$ \\
\texttt{cancer202030.0} & $[1023, 1481]$ & $40.546$ \\
\texttt{cancer202035.0} & $[1275, 1846]$ & $50.309$ \\
\texttt{cancer202040.0} & $[1372, 1986]$ & $54.393$ \\
\texttt{cancer202045.0} & $[1263, 1829]$ & $50.450$ \\
\texttt{cancer202050.0} & $[1090, 1579]$ & $44.376$ \\
\texttt{cancer202055.0} & $[934, 1352]$ & $38.388$ \\
\texttt{cancer202060.0} & $[674, 1234]$ & $32.982$ \\
\texttt{cancer202065.0} & $[443, 1002]$ & $25.461$ \\
\texttt{cancer202070.0} & $[278, 761]$ & $18.679$ \\
\texttt{cancer202075.0} & $[155, 505]$ & $12.085$ \\
\texttt{cancer202080.0} & $[62, 239]$ & $5.659$ \\
\texttt{cancer\_16} & $[0.45, 0.611]$ & $0.024$ \\
\texttt{cancer\_18} & $[0.098, 0.214]$ & $0.011$ \\
\texttt{cancer\_hi5} & $[0.104, 0.222]$ & $0.018$ \\
\texttt{cancer\_ohr} & $[0.115, 0.238]$ & $0.011$ \\
\texttt{cin3\_16} & $[0.173, 0.425]$ & $0.011$ \\
\texttt{cin3\_18} & $[0.034, 0.203]$ & $0.011$ \\
\texttt{cin3\_hi5} & $[0.227, 0.425]$ & $0.013$ \\
\texttt{cin3\_ohr} & $[0.173, 0.425]$ & $0.014$ \\
\hline
\end{tabular}
\end{center}

\end{document}